\documentclass[conference]{IEEEtran}
\IEEEoverridecommandlockouts
\usepackage{cite}
\usepackage{amsmath,amssymb,amsfonts}
\usepackage{algorithmic}
\usepackage{graphicx}
\usepackage{textcomp}
\usepackage{xcolor}

\usepackage[utf8]{inputenc} %
\usepackage[T1]{fontenc}    %
\usepackage{hyperref}       %
\usepackage{url}            %
\usepackage{booktabs}       %
\usepackage{amsfonts}       %
\usepackage{nicefrac}       %
\usepackage{microtype}      %

\usepackage{algorithmic}
\usepackage{graphicx}
\usepackage{textcomp}
\usepackage{xcolor}
\usepackage{multirow}
\usepackage{algorithm}
\usepackage{bm}
\usepackage{subfigure}
\usepackage{caption}
\usepackage{amsthm}

\newtheorem{theorem}{Theorem}
\newtheorem{assumption}{Assumption}

\def\BibTeX{{\rm B\kern-.05em{\sc i\kern-.025em b}\kern-.08em
    T\kern-.1667em\lower.7ex\hbox{E}\kern-.125emX}}
\begin{document}

\title{Beyond Double Ascent via Recurrent Neural Tangent Kernel in Sequential Recommendation}

\author{\IEEEauthorblockN{Ruihong Qiu, Zi Huang, and Hongzhi Yin\IEEEauthorrefmark{1}\thanks{* Corresponding author.}}
\IEEEauthorblockA{\textit{School of Information Technology and Electrical Engineering, The University of Queensland,}\\
r.qiu@uq.edu.au, huang@itee.uq.edu.au, h.yin1@uq.edu.au}
}

\maketitle

\begin{abstract}
Overfitting has long been considered a common issue to large neural network models in sequential recommendation. In our study, an interesting phenomenon is observed that overfitting is temporary. When the model scale is increased, the trend of the performance firstly ascends, then descends (i.e., overfitting) and finally ascends again, which is named as \textit{double ascent} in this paper. We therefore raise an assumption that a considerably larger model will generalise better with a higher performance. In an extreme case to infinite-width, performance is expected to reach the limit of this specific structure. Unfortunately, it is impractical to directly build a huge model due to the limit of resources. In this paper, we propose the Overparameterised Recommender (OverRec), which utilises a recurrent neural tangent kernel (RNTK) as a similarity measurement for user sequences to successfully bypass the restriction of hardware for huge models. We further prove that the RNTK for the tied input-output embeddings in recommendation is the same as the RNTK for general untied input-output embeddings, which makes RNTK theoretically suitable for recommendation. Since the RNTK is analytically derived, OverRec does not require any training, avoiding physically building the huge model. Extensive experiments are conducted on four datasets, which verifies the state-of-the-art performance of OverRec.\footnote{Code is available in \url{https://github.com/RuihongQiu/OverRec}}
\end{abstract}


\begin{figure}[!h]
    \centering
    \subfigure[Beauty.]{
    \includegraphics[width=0.39\linewidth]{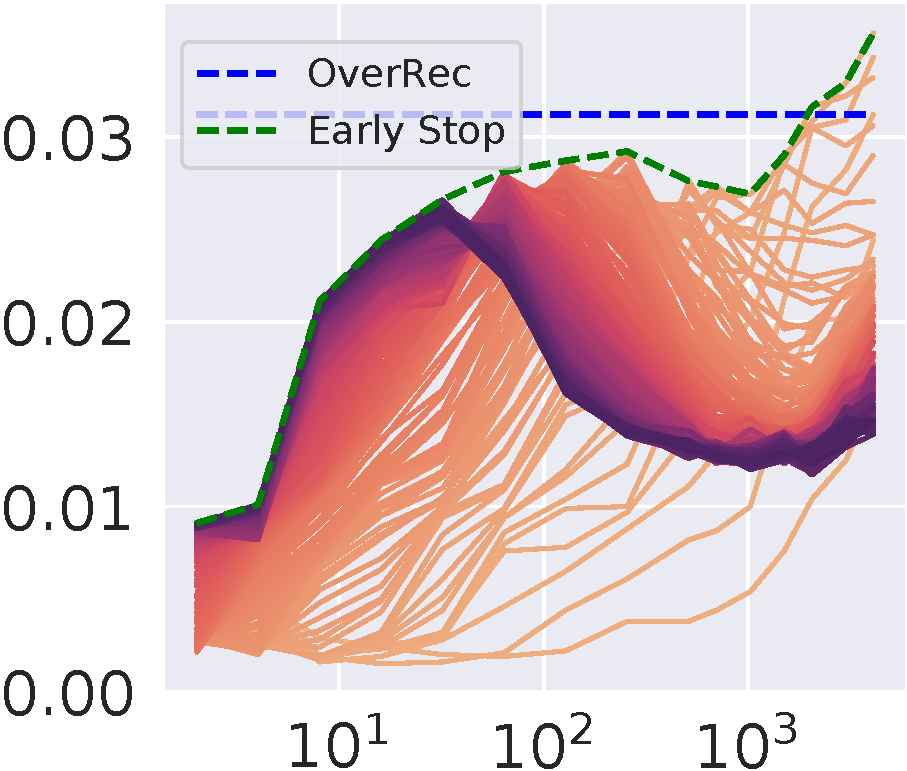}
    }
    \subfigure[Clothing.]{
    \includegraphics[width=0.52\linewidth]{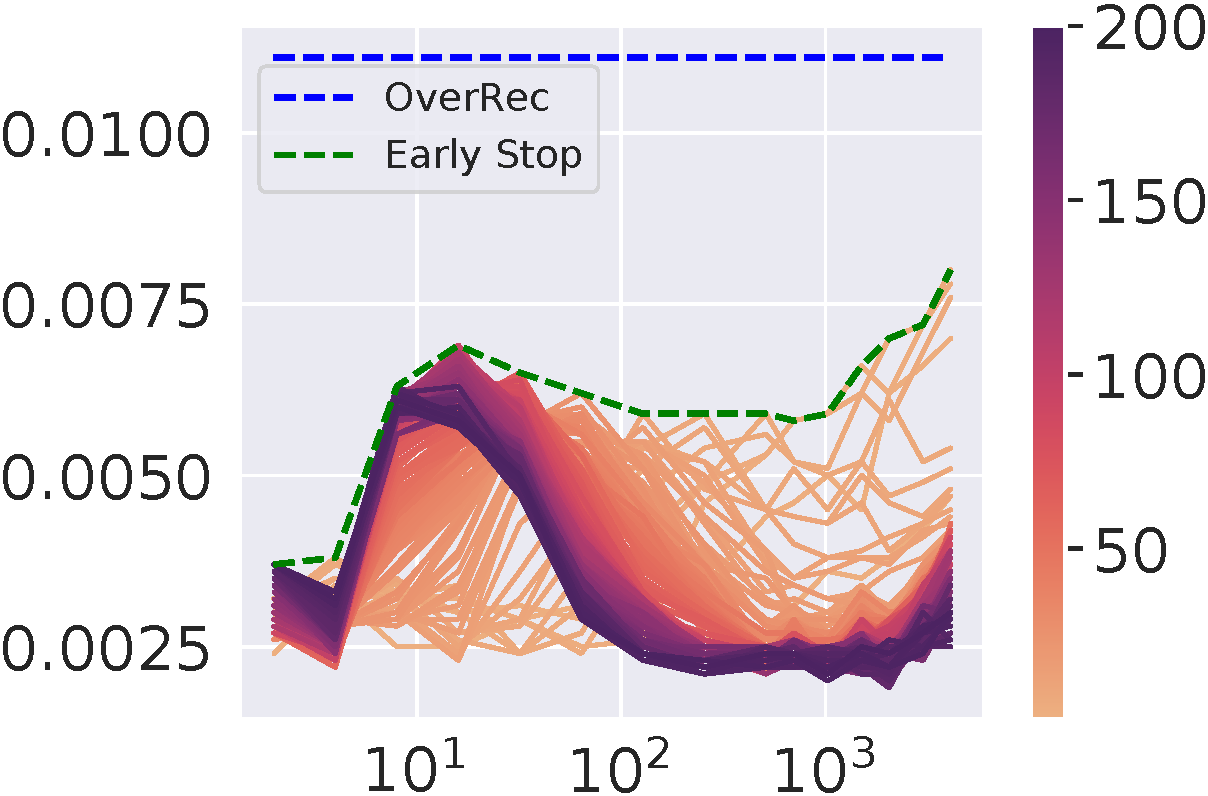}
    }
    \subfigure[Toys.]{
    \includegraphics[width=0.40\linewidth]{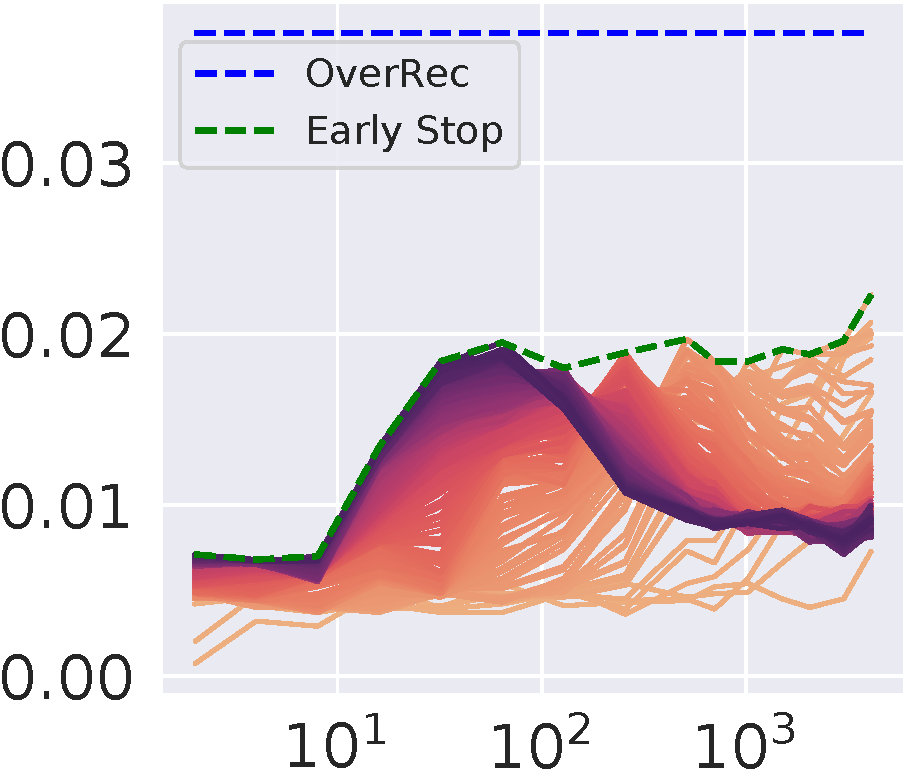}
    }
    \subfigure[ML-1M.]{
    \includegraphics[width=0.5\linewidth]{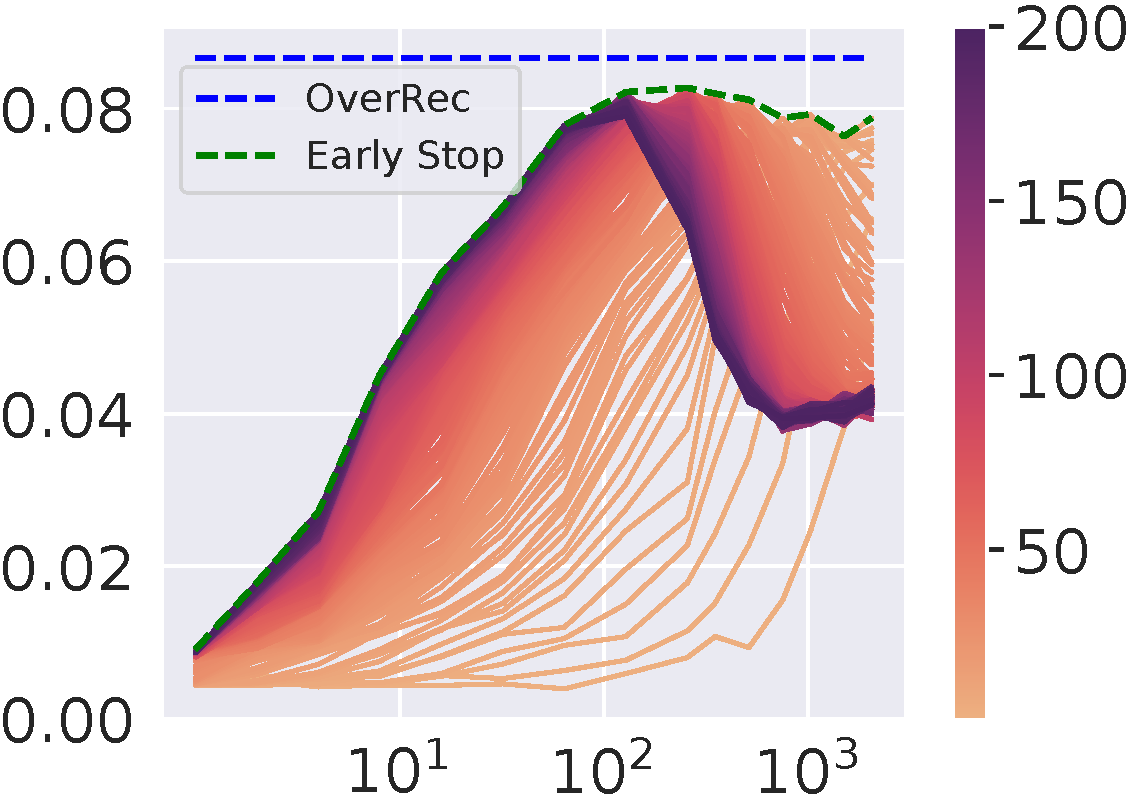}
    }
    \caption{\textit{Double ascent} on four datasets. X-axis is the width of models ranging from $2$ to $4096$. Y-axis is the NDCG@10. The colour bar denotes the training epoch from $1$ to $200$. Dots of the same horizontal value come from the same model. The blue dashed lines indicate the performance of OverRec. The green dashed lines indicate the highest performance of each model. (1) From green dashed lines, the test performance firstly ascends, then descends and finally ascends again with increasing the model size, which is named as the \textit{double ascent}. (2) OverRec can perform closely to the infinite-width RNN.}
    \label{fig:overpara}
\end{figure}

\section{Introduction}
Traditional recommender systems generally recommends content based on users' historical interactions without using the time factor~\cite{bprmf,fpmc,item-knn} although a user's preference often shifts. Recently, sequential recommendations have shown potentials to capture the dynamic preference~\cite{gru4rec,caser,sasrec,bert4rec,s3rec}.

In most of the empirical evidence of existing sequential models, overfitting seems to be inevitable for neural network models~\cite{sasrec,bert4rec,s3rec}. When the model scale, or alternatively, the number of parameters, grows to a certain extent, the overfitting will easily appear. Under this situation, though the training objective function will be optimised better, the generalisation of the trained model and the test performance will become worse. Thus, most existing methods use a relatively smaller model for higher performance~\cite{sasrec,bert4rec,s3rec}.

Interestingly, a \textit{double ascent} phenomenon different from overfitting is observed in our study that when further increasing the model scale. An empirical study using the classical recurrent neural networks (RNN)~\cite{rnn1,rnn2} is conducted on four datasets, Amazon Beauty (short form Beauty), Amazon Clothing, Shoes and Jewellery (short form Clothing), Amazon Toys and Games (short form Toys)~\cite{amazon} and MovieLens 1M (short form ML-1M)~\cite{movielens}. In this study, models using the same RNN paradigm are set to different widths, which are indicated by the size of the hidden dimension. Limited to our hardware configuration using NVIDIA V100 GPU, the choice of the width is from small values such as $2$ or $4$ to an overparameterisation regime (i.e., training a neural network with the number of parameters larger than the number of training data points) such as $3000$ or $4096$. In Fig.~\ref{fig:overpara}, as the model size increases, the performance of the Normalised Discounted Cumulative Gain (NDCG) @10 roughly will first ascend, then descend (traditionally considered as overfitting) and finally ascend again. This special trend of the performance is named as the \textit{double ascent}. We hypothesise that with a more powerful hardware, this trend is likely to continue and the performance will asymptotically converge to the limit of the specific model structure. To achieve a performance close to this limit, a model of very large scale is desired.

In real world, passively enlarging the model size, or even developing an infinite-width model, is impractical due to the limit of computational resources. Recent advances in the overparameterisation regime build a theoretical bridge between infinite-width neural networks and their corresponding kernels such as neural network Gaussian process kernel (NNGP)~\cite{nngp,nngp1,nngp2} and neural tangent kernel (NTK)~\cite{ntk,ntk-wide,ntk-exact}. These kernel theories state that under some mild conditions, the output of an infinite-width neural network at initialisation converges to NNGP, and the training evolution of the model is governed by NTK. With these kernels, the physical limitation of building a huge neural network can be resolved by its corresponding kernels, which is much more computational-friendly.

In this paper, we propose the Overparameterised Recommender (OverRec) model, which utilises a recurrent neural tangent kernel (RNTK) as the similarity measurement for user sequences to bypass the restriction of hardware for huge models. This RNTK is theoretically governing the evolution of the training of the infinite-width RNN, which retains the properties of infinite-width RNN. We further prove that with a mild independence assumption, the RNTK for RNN used in recommendation with tied input-output embeddings equals to the RNTK for RNN in general with untied input-output embeddings. Therefore, it is theoretically safe to use RNTK for recommendation and the prediction result of OverRec will naturally be similar to the result of the infinite-width recurrent model. Because the RNTK can be derived analytically, OverRec does not require any training. The main contributions of this paper are summarised as follows:
\begin{itemize}
    \item In overparameterisation regime of sequential recommendations, \textit{double ascent} is observed, which is different from the traditional overfitting when increasing the model size.
    \item The equivalence of RNTK between tied and untied input-output embedding situations is theoretically proved, which provides justifications to use RNTK in recommendations.
    \item With practical extensions over RNTK, a novel recommender system, OverRec, is proposed to obtain the power of infinite-width network in sequential recommendations.
    \item Extensive experiments are conducted on four benchmark datasets, which verifies the effectiveness to mimic infinite RNN and the superiority of OverRec.
\end{itemize}

\section{Preliminaries}
\subsection{Infinite-Width Fully-connected Neural Networks and Kernels}
In the following, bold lowercase and uppercase symbols are used to denote vectors and matrices respectively. Superscripts are used for indexing different vectors or matrices and subscripts are used for indexing different elements within vectors or matrices. For a fully-connected neural network $f_{\boldsymbol{\theta}}(\boldsymbol{x})$ with $\boldsymbol{\theta}$ as parameters and $\boldsymbol{x}$ as input, the infinite-width indicates that the number of weights in hidden layers of $f$ goes to infinity. If this infinite network is initialised with $\boldsymbol{\theta}$ drawn independently and identically distributed (i.i.d.) from zero-mean Gaussian distributions, the outputs of the network with different inputs converge to a Gaussian process with the kernel:
\begin{equation}
\label{eq:nngp}
    \mathcal{K}(\boldsymbol{x},\boldsymbol{x}')=\mathbb{E}\left[f_{\boldsymbol{\theta}_0,i}(\boldsymbol{x})\cdot f_{\boldsymbol{\theta}_0,i}(\boldsymbol{x}')\right],
\end{equation}
which denotes the covariance between the $i$-th dimension in the outputs of input $\boldsymbol{x}$ and input $\boldsymbol{x}'$ respectively fed into a network with random parameters $\boldsymbol{\theta}_0$. Therefore, the output of the infinite network with multiple inputs, $\mathcal{X}$, at initialisation is a Gaussian process, $f_{\boldsymbol{\theta}_0}(\mathcal{X})\sim\mathcal{N}(0,\mathcal{K}(\mathcal{X},\mathcal{X}))$, called neural network Gaussian process (NNGP). For most structures in deep learning, the NNGP $\mathcal{K}$ can be computed analytically~\cite{nngp,tp1gp4a}.

For infinite network, if it is trained with gradient descent, it can be approximated by a linear model~\cite{ntk-wide,lazy}:
\begin{equation}
\label{eq:lin}
    f_{\boldsymbol{\theta}_a}^\text{lin}\approx f_{\boldsymbol{\theta}_0}(\boldsymbol{x})+\left\langle\nabla_{\boldsymbol{\theta}_a} f_{\boldsymbol{\theta}_0}(\boldsymbol{x}), \boldsymbol{\theta}_a-\boldsymbol{\theta}_0\right\rangle,
\end{equation}
where $a$ denotes the $a$-th step of gradient descent. The training of infinite network is equivalent to kernel gradient descent with a neural tangent kernel (NTK) from the derivatives:
\begin{equation}
\label{eq:ntk}
    \Theta_a\left(\boldsymbol{x}, \boldsymbol{x}^{\prime}\right)=\mathbb{E}\left[\nabla_{\boldsymbol{\theta}_a} f_{\boldsymbol{\theta}_a}(\boldsymbol{x})^{\mathrm{T}}\cdot\nabla_{\boldsymbol{\theta}_a} f_{\boldsymbol{\theta}_a}\left(\boldsymbol{x}^{\prime}\right)\right].
\end{equation}

Recent work proved that the error of the approximation converges to zero as the width goes to infinity~\cite{ntk,ntk-wide,lazy}. More importantly, the NTK nearly stay the same as initialisation after training, which is called lazy training~\cite{ntk,ntk-wide,lazy}:
\begin{equation}
    \lim _{n \rightarrow \infty} \Theta_{a}\left(\boldsymbol{x}, \boldsymbol{x}^{\prime}\right)=\lim _{n \rightarrow \infty} \Theta_{0}\left(\boldsymbol{x}, \boldsymbol{x}^{\prime}\right):=\Theta\left(\boldsymbol{x}, \boldsymbol{x}^{\prime}\right), \quad \forall a,
\end{equation}
where $n$ is the width of the network. It is safe to use $\Theta_0$ at initialisation to approximate the well-trained infinite network.

\subsection{Recurrent Neural Networks}
For an input sequence $\boldsymbol{x}=\left[\boldsymbol{x}_t\right]^T_{t=1}$ of length $T$, where $\boldsymbol{x}_t\in\mathbb{R}^m$, a recurrent neural network (RNN)~\cite{rnn1,rnn2} with $L$ layers will compute as follows at time step $t$ for layer $l\in L$:
\begin{align}
\label{eq:rnn}
    \boldsymbol{g}^{(l, t)}(\boldsymbol{x})&=\boldsymbol{W}^{(l)} \boldsymbol{h}^{(l, t-1)}(\boldsymbol{x})+\boldsymbol{U}^{(l)} \boldsymbol{h}^{(l-1, t)}(\boldsymbol{x})+\boldsymbol{b}^{(l)},\notag\\
    \boldsymbol{h}^{(l, t)}(\boldsymbol{x})&=\phi\left(\boldsymbol{g}^{(l, t)}(\boldsymbol{x})\right),
\end{align}
where $\boldsymbol{g}^{(l, t)}$ is the pre-activation vector at time step $t$ for layer $l$ and $\boldsymbol{h}^{(l, t)}$ is the hidden state after applying an activation function $\phi$. $\boldsymbol{W}^{(l)}\in\mathbb{R}^{n\times n}$, $\boldsymbol{b}^{(l)}\in\mathbb{R}^n$, $\boldsymbol{U}^{(1)}\in\mathbb{R}^{n\times m}$ and $\boldsymbol{U}^{(l)}\in\mathbb{R}^{n\times n},l\in[2,L],$ are the trainable parameters of RNN. For input layer $l=0$, $\boldsymbol{h}^{(0, t)}(\boldsymbol{x})=\boldsymbol{x}_t$.

In the output layer of RNN, a linear function is used as:
\begin{equation}
\label{eq:rnnoutput}
    f^{(t)}_{\boldsymbol{\theta}}(\boldsymbol{x})=\boldsymbol{V}\boldsymbol{h}^{(L, t)}(\boldsymbol{x}),
\end{equation}
where $\boldsymbol{V}\in\mathbb{R}^{d\times n}$ is the output weight.

\section{Recurrent Neural Tangent Kernel}

\subsection{Initialisation for Recurrent Neural Tangent Kernel}
Following recent work~\cite{ntk,rntk}, the parameter initialisation for infinite-width RNN is defined as follows:
\begin{align}
\label{eq:init}
    \boldsymbol{W}^{(l)}&=\frac{\sigma_{W}^{l}}{\sqrt{n}} \mathbf{W}^{(l)},\quad\boldsymbol{b}^{(l)}=\sigma_{b} \mathbf{b}^{(l)},\quad\boldsymbol{V}=\frac{\sigma_{V}}{\sqrt{n}} \mathbf{V},\notag\\ \boldsymbol{U}^{(1)}&=\frac{\sigma_{U}^{1}}{\sqrt{m}} \mathbf{U}^{(1)},\quad\boldsymbol{U}^{(l)}=\frac{\sigma_{U}^{l}}{\sqrt{n}} \mathbf{U}^{(l)}(l \geq 2),
\end{align}
where $\mathbf{W}^{(l)}_{i,j},\mathbf{b}^{(l)}_i,\mathbf{V}_{i,j},\mathbf{U}^{(l)}_{i,j}\sim\mathcal{N}(0,1)$. This initialisation is called NTK initialisation. The choice of $\sigma_{W},\sigma_{b},\sigma_{U}$ and $\sigma_{V}$ controls the variance of the parameters at initialisation of RNN. The initial hidden state $\boldsymbol{h}^{(l,0)}$ is set to zero. To have a consistent notation, all learnable parameters are wrapped as $\boldsymbol{\theta}=\left[\left\{\mathbf{W}^{(l)},\mathbf{b}^{(l)},\mathbf{U}^{(l)}\right\}_{l=1}^L,\mathbf{V}\right]$.

\subsection{Neural Network Gaussian Process for Recurrent Neural Networks}
\label{sec:nngp-rnn}
The kernel of the neural network Gaussian process for recurrent neural networks (NNGP-RNN) describes the behaviour of output. Firstly, as the width of the network goes to infinity, every dimension of the pre-activation $\boldsymbol{g}^{(l, t)}(\boldsymbol{x})$ converges to a zero-mean Gaussian process with the following kernel:
\begin{equation}
\label{eq:kernel-pre}
    \Sigma^{(l,t,t')}(\boldsymbol{x},\boldsymbol{x}')=\underset{\boldsymbol{\theta} \sim \mathcal{N}(0,1)}{\mathbb{E}}\left[\boldsymbol{g}^{(l, t)}_i(\boldsymbol{x})\cdot\boldsymbol{g}^{(l, t')}_i(\boldsymbol{x}')\right].
\end{equation}
This convergence exists because the parameters $\left\{\boldsymbol{W},\boldsymbol{U},\boldsymbol{b}\right\}$ in Equation (\ref{eq:rnn}) are i.i.d. based on Equation (\ref{eq:init}) and when the width goes to infinity, the summation of the associated terms of i.i.d. parameters will converge to Gaussian according to the law of large numbers and the central limit theorem.

From the definition of NNGP, it considers the outputs of to different inputs and converges to another Gaussian process. NNGP-RNN is defined similarly to Equation (\ref{eq:nngp}):
\begin{equation}
\label{eq:nngp-rnn}
    \mathcal{K}(\boldsymbol{x},\boldsymbol{x}')=\underset{\boldsymbol{\theta} \sim \mathcal{N}(0,1)}{\mathbb{E}}\left[f_{\boldsymbol{\theta},i}(\boldsymbol{x}) \cdot f_{\boldsymbol{\theta},i}\left(\boldsymbol{x}^{\prime}\right)\right].
\end{equation}
This convergence exists similarly to Equation (\ref{eq:kernel-pre}). Although there is a nonlinear function $\phi$, the parameters of $\boldsymbol{V}$ are still i.i.d. and the central limit theorem results in a Gaussian distribution.

\begin{algorithm}[t]
    \caption{Analytic NNGP-RNN for same length sequences}
    \label{alg:nngp-rnn}
    \begin{algorithmic}[1]
    \REQUIRE $\boldsymbol{x},\boldsymbol{x}',\sigma_{W},\sigma_{b},\sigma_{U},\sigma_{V},\phi,L$;\
    \ENSURE $\mathcal{K}^{(T)}(\boldsymbol{x},\boldsymbol{x}')$;\
    \STATE $\Sigma^{(1,1)}(\boldsymbol{x},\boldsymbol{x}')=\sigma^2_U\left[\boldsymbol{x}^{(1)}\cdot{\boldsymbol{x}'}^{(1)}\right]+\sigma^2_b$;\label{alg:nngp-rnn-dot1}
    \FOR{$l=2,\dots,L$}
    \STATE $\Sigma^{(l,1)}(\boldsymbol{x},\boldsymbol{x}')=\sigma^2_UV_\phi\left[\boldsymbol{K}^{(l-1,1)}(\boldsymbol{x},{\boldsymbol{x}'})\right]+\sigma^2_b$;
    \ENDFOR
    \FOR{$t=2,\dots,T$}
    \STATE $\begin{aligned}\Sigma^{(1,t)}(\boldsymbol{x},\boldsymbol{x}')&=\sigma^2_U\left[\boldsymbol{x}^{(t)}\cdot{\boldsymbol{x}'}^{(t)}\right]\notag\\
    &+\sigma_W^2V_\phi\left[\boldsymbol{K}^{(1,t-1)}(\boldsymbol{x},{\boldsymbol{x}'})\right]+\sigma^2_b;\end{aligned}$\label{alg:nngp-rnn-dot2}
    \FOR{$l=2,\dots,L$}
    \STATE $\begin{aligned}\Sigma^{(l,t)}(\boldsymbol{x},\boldsymbol{x}')&=\sigma^2_UV_\phi\left[\boldsymbol{K}^{(l-1,t)}(\boldsymbol{x},{\boldsymbol{x}'})\right]\notag\\
        &+\sigma_W^2V_\phi\left[\boldsymbol{K}^{(l,t-1)}(\boldsymbol{x},{\boldsymbol{x}'})\right]+\sigma^2_b;\notag\end{aligned}$
    \ENDFOR
    \ENDFOR
    \STATE $\mathcal{K}^{(T)}(\boldsymbol{x},\boldsymbol{x}')=\sigma^2_VV_\phi\left[\boldsymbol{K}^{(L,T)}(\boldsymbol{x},{\boldsymbol{x}'})\right]$;
    \end{algorithmic}
\end{algorithm}

NNGP-RNN can be calculated following~\cite{tp1gp4a,rntk}. The calculation for two sequences with the same length is in Algorithm~\ref{alg:nngp-rnn}. Notice that if there is nonlinear function involved, $V_\phi\left[\boldsymbol{K}\right]$ is an operator based on nonlinear activation function $\phi$ and a positive definite matrix $\boldsymbol{K}\in\mathbb{R}^{2\times2}$:
\begin{equation}
    V_\phi\left[\boldsymbol{K}\right]=\underset{\boldsymbol{\theta} \sim \mathcal{N}(0,1)}{\mathbb{E}}\left[\phi(\boldsymbol{g}_i)\cdot\phi({\boldsymbol{g}'}_i)\right],\quad(\boldsymbol{g}_i,{\boldsymbol{g}'}_i)\sim\mathcal{N}(0,\boldsymbol{K}).
\end{equation}
Generally, $\phi$ can be Rectified Linear Unit (ReLU)~\cite{relu}. The analytical forms of $V_\phi$ given $\boldsymbol{K}=\left[\begin{array}{ll}
K_{1} & K_{3} \\
K_{3} & K_{2}
\end{array}\right]$ are~\cite{relu-v}:
\begin{equation}
    V_{\text{ReLU}}[\boldsymbol{K}]=\frac{1}{2 \pi}\left(c(\pi-\arccos (c))+\sqrt{1-c^{2}}\right) \sqrt{K_{1} K_{2}},
\end{equation}
where $c=K_3/\sqrt{K_1K_2}$. In Algorithm~\ref{alg:nngp-rnn}, $\boldsymbol{K}$ is defined as:
\begin{equation}
\label{eq:k}
    \boldsymbol{K}^{(l,t)}(\boldsymbol{x},\boldsymbol{x}')=\left[\begin{array}{ll}
\Sigma^{(l,t)}(\boldsymbol{x},\boldsymbol{x}) & \Sigma^{(l,t)}(\boldsymbol{x},\boldsymbol{x}') \\
\Sigma^{(l,t)}(\boldsymbol{x}',\boldsymbol{x}) & \Sigma^{(l,t)}(\boldsymbol{x}',\boldsymbol{x}')
\end{array}\right].
\end{equation}

\subsection{Recurrent Neural Tangent Kernel}
\label{sec:rntk}
The RNTK describes the derivatives against the parameters $\boldsymbol{\theta}$ of different inputs. According to~\cite{tp2ntk4a}, the gradients of a neural network during backpropagation are distributed as Gaussian. Define the gradient vector of $\boldsymbol{g}$ as $\boldsymbol{\delta}^{(l, t)}(\boldsymbol{x})=\sqrt{n}\left(\nabla_{\boldsymbol{g}^{(l, t)}(\boldsymbol{x})} f_{\boldsymbol{\theta}}(\boldsymbol{x})\right)$ and the corresponding kernel is:
\begin{equation}
    \Pi^{\left(\ell, t, t^{\prime}\right)}\left(\boldsymbol{x}, \boldsymbol{x}^{\prime}\right)=\underset{\theta \sim \mathcal{N}(0,1)}{\mathbb{E}}\left[\boldsymbol{\delta}_{i}^{(l, t)}(\boldsymbol{x}) \cdot\boldsymbol{\delta}_{i}^{\left(l, t^{\prime}\right)}\left(\boldsymbol{x}^{\prime}\right)\right].
\end{equation}

With the above kernels, the RNTK is theoretically proved:
\begin{theorem}
\label{thm:rntk}
\cite{rntk} Let $\boldsymbol{x}$ and $\boldsymbol{x}'$ be two sequences with lengths $T$ and $T'$. Without loss of generality, assume that $T\leq T'$ and let $\tau=T'-T$. The RNTK for an $L$ layer infinite RNN is:
\begin{align}
\label{eq:rntk}
    &\Theta^{\left(L, T, T^{\prime}\right)}\left(\boldsymbol{x}, \boldsymbol{x}^{\prime}\right)=\left\langle\nabla_{\boldsymbol{\theta}_t} f_{\boldsymbol{\theta}_t}(\boldsymbol{x}), \nabla_{\boldsymbol{\theta}_t} f_{\boldsymbol{\theta}_t}\left(\boldsymbol{x}^{\prime}\right)\right\rangle=\notag\\
    &\left(\sum_{l=1}^{L} \sum_{t=1}^{T}\left(\Pi^{(l, t, t+\tau)}\left(\boldsymbol{x}, \boldsymbol{x}^{\prime}\right) \cdot \Sigma^{(l, t, t+\tau)}\left(\boldsymbol{x}, \boldsymbol{x}^{\prime}\right)\right)+\mathcal{K}\left(\boldsymbol{x}, \boldsymbol{x}^{\prime}\right)\right)\notag\\
    &\otimes\boldsymbol{I},
\end{align}
where $\otimes$ is element-wise multiplication and $\boldsymbol{I}$ is identity matrix.
\end{theorem}

From Theorem~\ref{thm:rntk} and Equation (\ref{eq:lin}), it can be concluded that with a well-trained infinite-width RNN $f$, the approximated linearised model $f^{\text{lin}}$ is governed by the RNTK $\Theta$. The convergence rate of the approximation is $\mathcal{O}(n^{-\frac{1}{2}})$~\cite{ntk-wide}.

\begin{algorithm}[t]
    \caption{Analytic RNTK for same length sequences}
    \label{alg:rntk}
    \begin{algorithmic}[1]
    \REQUIRE $\boldsymbol{x},\boldsymbol{x}',\sigma_{W},\sigma_{b},\sigma_{U},\sigma_{V},\phi,L,\Sigma(\boldsymbol{x},\boldsymbol{x}'),\mathcal{K}(\boldsymbol{x},\boldsymbol{x}')$;\
    \ENSURE $\Theta^{(T)}(\boldsymbol{x},\boldsymbol{x}')$;\
    \STATE $\Psi^{(1,1)}(\boldsymbol{x},\boldsymbol{x}')=\Sigma^{(1,1)}(\boldsymbol{x},\boldsymbol{x}')$;
    \FOR{$l=2,\dots,L$}
    \STATE $\begin{aligned}\Psi^{(l,1)}&(\boldsymbol{x},\boldsymbol{x}')=\Sigma^{(l,1)}(\boldsymbol{x},\boldsymbol{x}')\notag\\
    &+\sigma^2_U\Psi^{(l-1,1)}(\boldsymbol{x},\boldsymbol{x}')V_{\phi'}\left[\boldsymbol{K}^{(l-1,1)}(\boldsymbol{x},{\boldsymbol{x}'})\right];\notag\end{aligned}$
    \ENDFOR
    \FOR{$t=2,\dots,T$}
    \STATE $\begin{aligned}\Psi^{(1,t)}&(\boldsymbol{x},\boldsymbol{x}')=\Sigma^{(1,t)}(\boldsymbol{x},\boldsymbol{x}')\notag\\
    &+\sigma^2_W\Psi^{(1,t-1)}(\boldsymbol{x},\boldsymbol{x}')V_{\phi'}\left[\boldsymbol{K}^{(1,t-1)}(\boldsymbol{x},{\boldsymbol{x}'})\right];\notag\end{aligned}$
    \FOR{$l=2,\dots,L$}
    \STATE $\begin{aligned}\Psi^{(l,t)}&(\boldsymbol{x},\boldsymbol{x}')=\Sigma^{(l,t)}(\boldsymbol{x},\boldsymbol{x}')\notag\\
    &+\sigma^2_W\Psi^{(l,t-1)}(\boldsymbol{x},\boldsymbol{x}')V_{\phi'}\left[\boldsymbol{K}^{(l,t-1)}(\boldsymbol{x},{\boldsymbol{x}'})\right]\notag\\
    &+\sigma^2_U\Psi^{(l-1,t)}(\boldsymbol{x},\boldsymbol{x}')V_{\phi'}\left[\boldsymbol{K}^{(l-1,t)}(\boldsymbol{x},{\boldsymbol{x}'})\right];\notag\end{aligned}$
    \ENDFOR
    \ENDFOR
    \STATE $\begin{aligned}\Theta^{(T)}&(\boldsymbol{x},\boldsymbol{x}')=\mathcal{K}^{(T)}(\boldsymbol{x},\boldsymbol{x}')\notag\\
    &+\sigma^2_U\Psi^{(L,T)}(\boldsymbol{x},\boldsymbol{x}')V_{\phi'}\left[\boldsymbol{K}^{(L,T)}(\boldsymbol{x},{\boldsymbol{x}'})\right];\end{aligned}$
    \end{algorithmic}
\end{algorithm}

The detailed RNTK calculation for two sequences with the same length is in Algorithm~\ref{alg:rntk}. $\Psi$ is an accumulation term for $\Theta$ and $\Pi$. $V_{\phi'}[\boldsymbol{K}]$ is an operator of derivatives for $\phi$ (ReLU) with the positive definite matrix $\boldsymbol{K}\in\mathbb{R}^{2\times2}$ in Equation (\ref{eq:k}):
\begin{equation}
    V_{\text{ReLU}'}[\boldsymbol{K}]=\frac{1}{2 \pi}(\pi-\arccos (c)).
\end{equation}

\subsection{RNTK with Tied Input-Output Embedding}
In general recommendation setting, the input-output embedding are tied. It means that the input layer weight matrix $\boldsymbol{U}^{(1)}$ in Equation (\ref{eq:rnn}) and the output layer weight matrix $\boldsymbol{V}$ in Equation (\ref{eq:rnnoutput}) are using the same embedding matrix. Mathematically, it can be viewed as $\boldsymbol{V}=\boldsymbol{U}^{(1),\mathrm{T}}$, where $\mathrm{T}$ means transpose.

However, the existing RNTK derivations from~\cite{rntk,tp2ntk4a} are all based on the general setting that the input-output embedding are untied. Mathematically, it can be viewed as $\boldsymbol{V}\neq\boldsymbol{U}^{(1),\mathrm{T}}$

In the derivation of NTK, there is a widely applied assumption, Gradient Independence Assumption~\cite{rntk,tp2ntk4a,dip}. This assumption describes that in infinite-width neural networks, a weight $\boldsymbol{W}$ used in forward computing and its transpose $\boldsymbol{W}^{\mathrm{T}}$ used in the backward computing are independent.

\begin{assumption}
\label{asp:gia}
(Gradient Independence Assumption)~\cite{rntk,tp2ntk4a,dip} In infinite-width neural networks, weights used in calculating gradients ($\boldsymbol{W}^{\mathrm{T}}$) is an i.i.d.\ copy of weights used in forward propagation ($\boldsymbol{W}$) if the output weight ($\boldsymbol{V}$) is sampled independently from other parameters and has mean 0.
\end{assumption}

This assumption can be understood in an intuitive way that the independently sampled output layer weight $\boldsymbol{V}$ with mean 0 blocks all potential correlation between $\boldsymbol{W}$ and $\boldsymbol{W}^{\mathrm{T}}$.

To deal with the tied input-output embedding situation in recommendation, we extend this assumption in the following way to Embedding Independence Assumption:

\begin{assumption}
\label{asp:eia}
(Embedding Independence Assumption) In infinite-width neural networks, an embedding matrix ($\boldsymbol{U}^{\mathrm{T}}$) is an i.i.d.\ copy of ($\boldsymbol{U}$) if the intermediate layer weight ($\boldsymbol{W}$) is sampled independently from other parameters and has mean 0, and there is no directly additive connections in between.
\end{assumption}

Under this assumption, the behaviour of $\boldsymbol{U}^{\mathrm{T}}$ in forward propagation mimics the behaviour of $\boldsymbol{W}^{\mathrm{T}}$ in backward propagation in Assumption~\ref{asp:gia}. In Assumption~\ref{asp:eia}, the intermediate layer weight can safely successfully block the correlation between the embedding matrix and its transpose. However, if there are directly additive connections between $\boldsymbol{U}$ and $\boldsymbol{U}^{\mathrm{T}}$, such as a residual connection, the independence assumption is violated. In an extreme case, an infinite-width Word2Vec~\cite{word2vec}, if a tied embedding matrix is used, the matrix and its transpose cannot be treated as independent in infinite-width situation because there is no intermediate layer in between.

With this extension of assumption, we can calculate the NTK for tied input-output embedding recurrent neural network:

\begin{theorem}
\label{thm:tied}
For the same infinite-width recurrent structure, the neural tangent kernel of untied input-output embedding recurrent neural network and the neural tangent kernel of tied input-output embedding recurrent neural network are the same, when the recurrent neural network has more than one layer or the length of the input sequence is longer than one.
\end{theorem}

The proof of this theorem is a straight-forward extension of Theorem~\ref{thm:rntk} under Assumption~\ref{asp:eia}. In the derivation of RNTK, there are terms that will be affected after substituting the output layer weight $\boldsymbol{V}$ with $\boldsymbol{U}^{(1),\mathrm{T}}$. These terms originally regard the covariance between $\boldsymbol{V}$ and all other weights~\cite{rntk}. After the substitution, all these covariance terms remain unchanged, except that the original covariance between $\boldsymbol{V}$ and $\boldsymbol{U}^{(1)}$, which becomes the covariance between $\boldsymbol{U}^{(1),\mathrm{T}}$ and $\boldsymbol{U}^{(1)}$. With Assumption~\ref{asp:eia}, the covariance between $\boldsymbol{U}^{(1),\mathrm{T}}$ and $\boldsymbol{U}^{(1)}$ is zero, which is the same as the covariance between $\boldsymbol{V}$ and $\boldsymbol{U}^{(1)}$. This extension is valid when the number of RNN layer and the length of the input sequence do not equal to one at the same time. This is because when this situation happens, there is a direct additive connection between $\boldsymbol{U}^{(1)}$ and $\boldsymbol{U}^{(1),\mathrm{T}}$ according to Equation (\ref{eq:rnn}) and (\ref{eq:rnnoutput}), which will violate Assumption~\ref{asp:eia}.

\begin{figure}[!t]
    \centering
    \subfigure[Beauty.]{
    \includegraphics[width=0.45\linewidth]{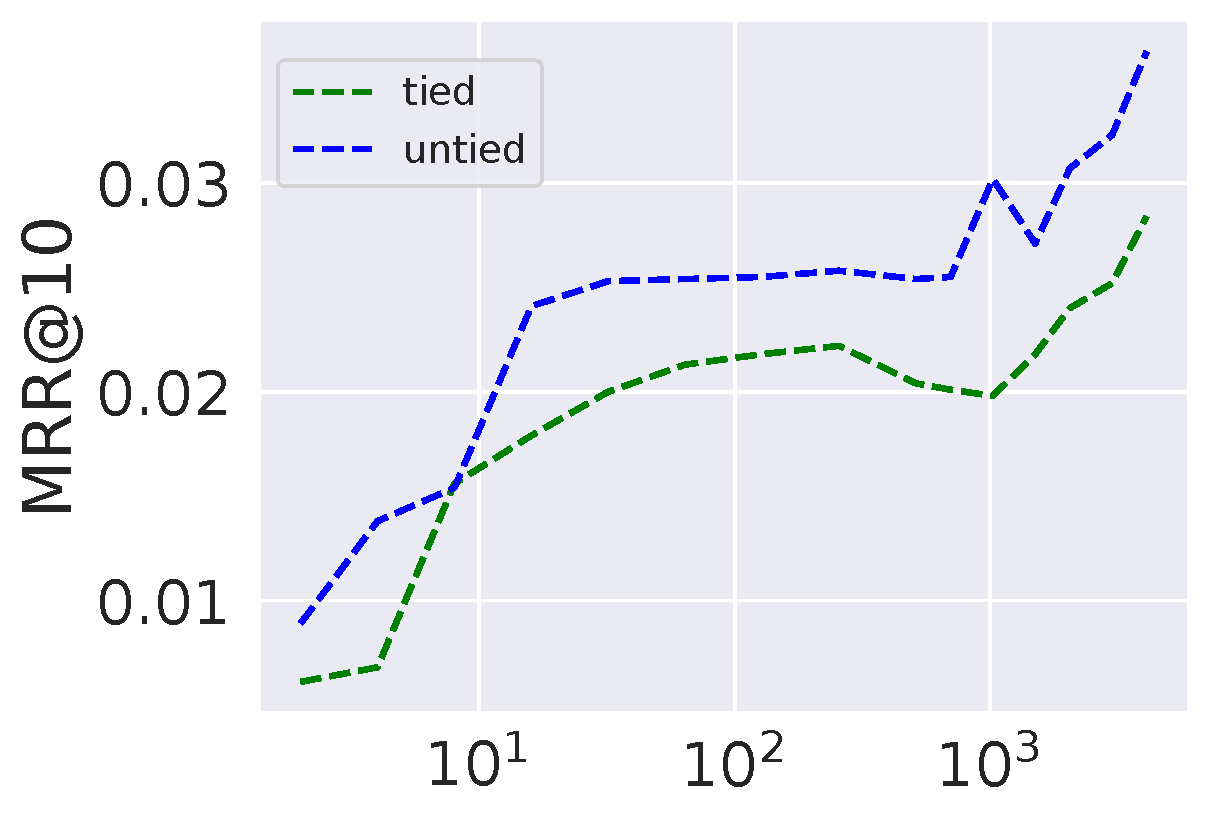}
    }
    \subfigure[Clothing.]{
    \includegraphics[width=0.47\linewidth]{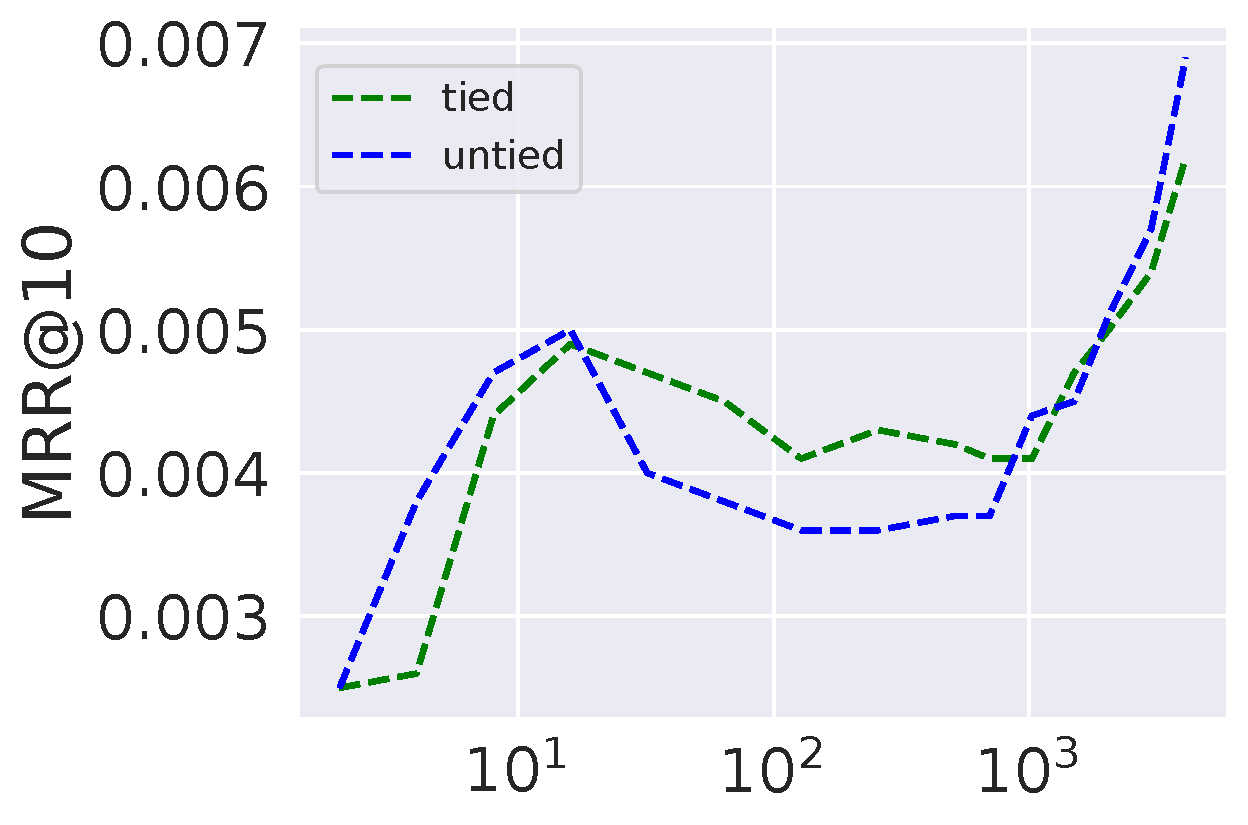}
    }
    \subfigure[Toys.]{
    \includegraphics[width=0.47\linewidth]{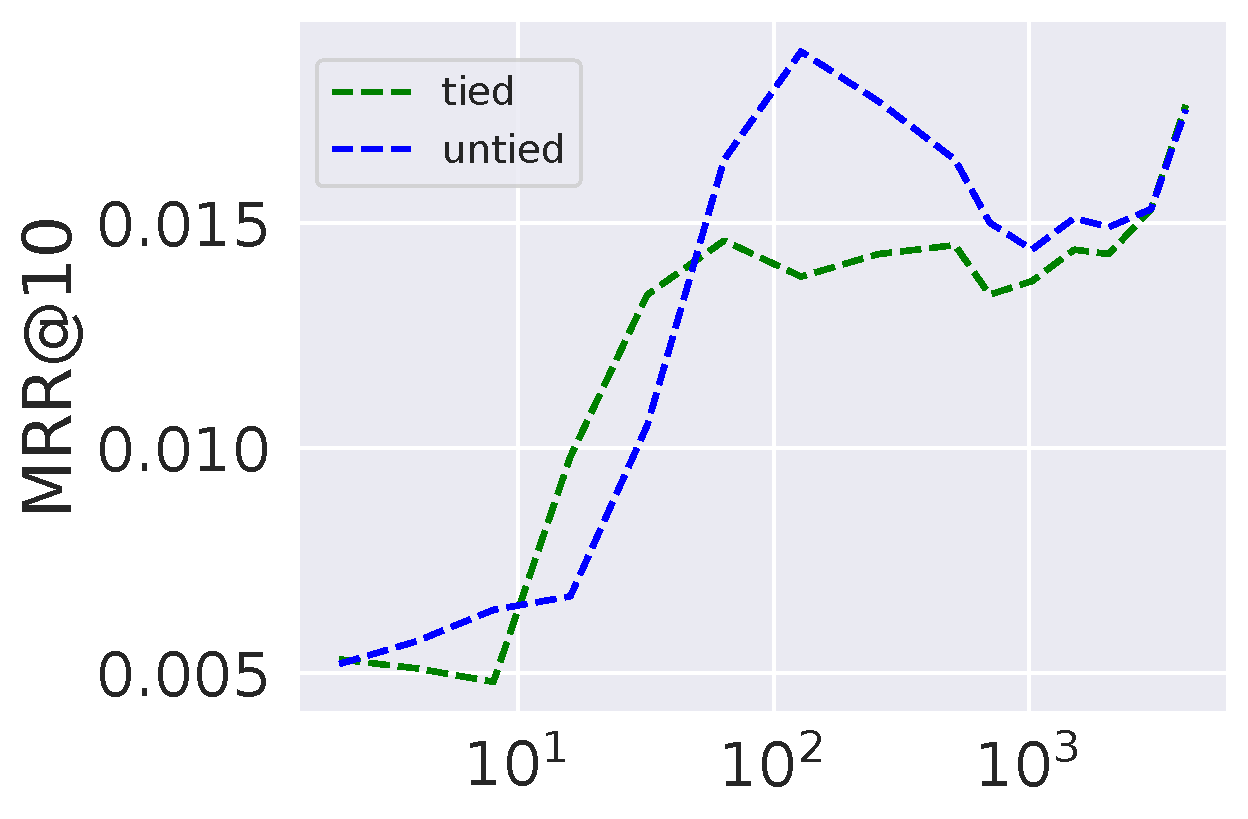}
    }
    \subfigure[ML-1M.]{
    \includegraphics[width=0.46\linewidth]{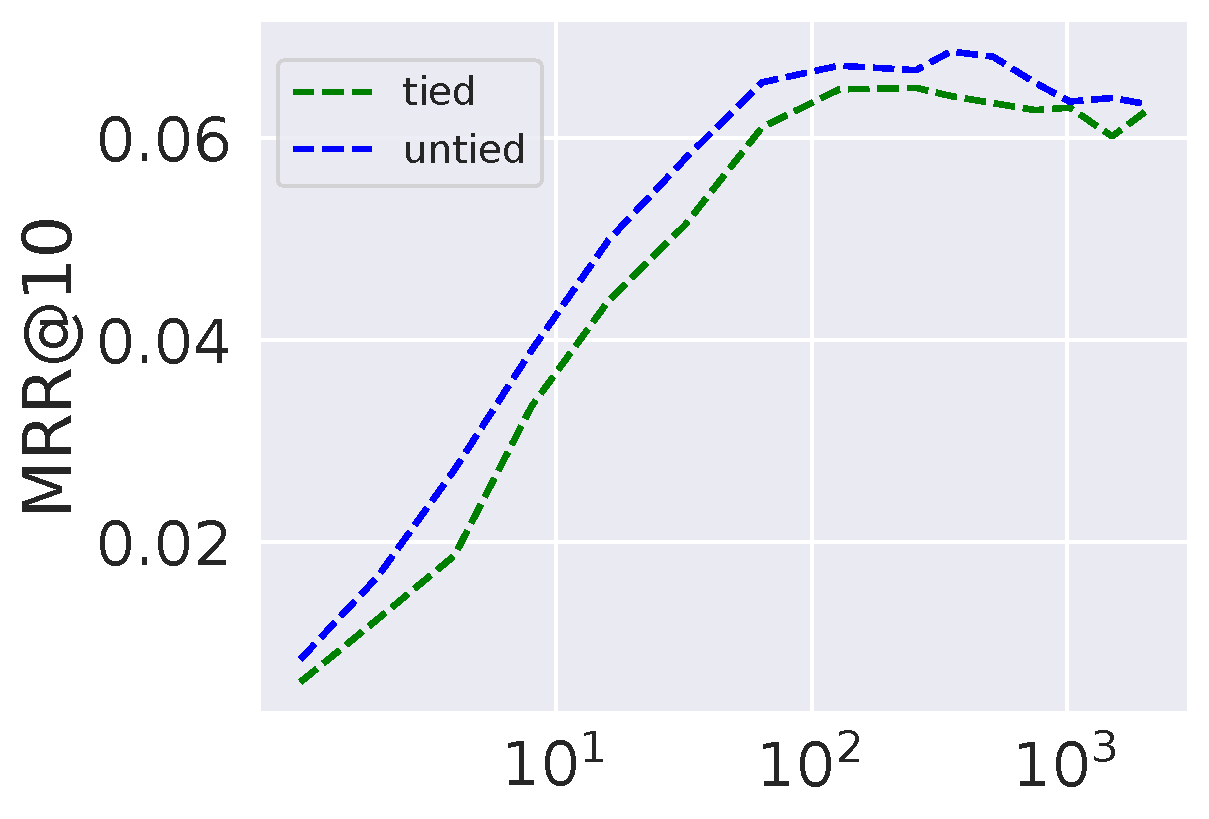}
    }
    \caption{Recommendation performance of RNN encoder with tied input-output embeddings (green) and untied input-output embeddings (blue) with respect to hidden dimension.}
\label{fig:tie}
\end{figure}

Thus, the derivation of RNTK will not change by tying the input-output embeddings. Furthermore, in the infinite-width situation of RNN, the theoretical performances of untied input-output embeddings and tied input-output embeddings are the same. The empirical performances of these two types of RNN on four datasets are shown in Fig.~\ref{fig:tie}. The performances of tied (green) and untied (blue) situations are close for large model size. Consequently, RNTK for RNN with untied input-output embeddings can be safely deployed in sequential recommendation where there are tied input-output embeddings.

\begin{figure}[t]
    \centering
    \includegraphics[width=0.7\linewidth]{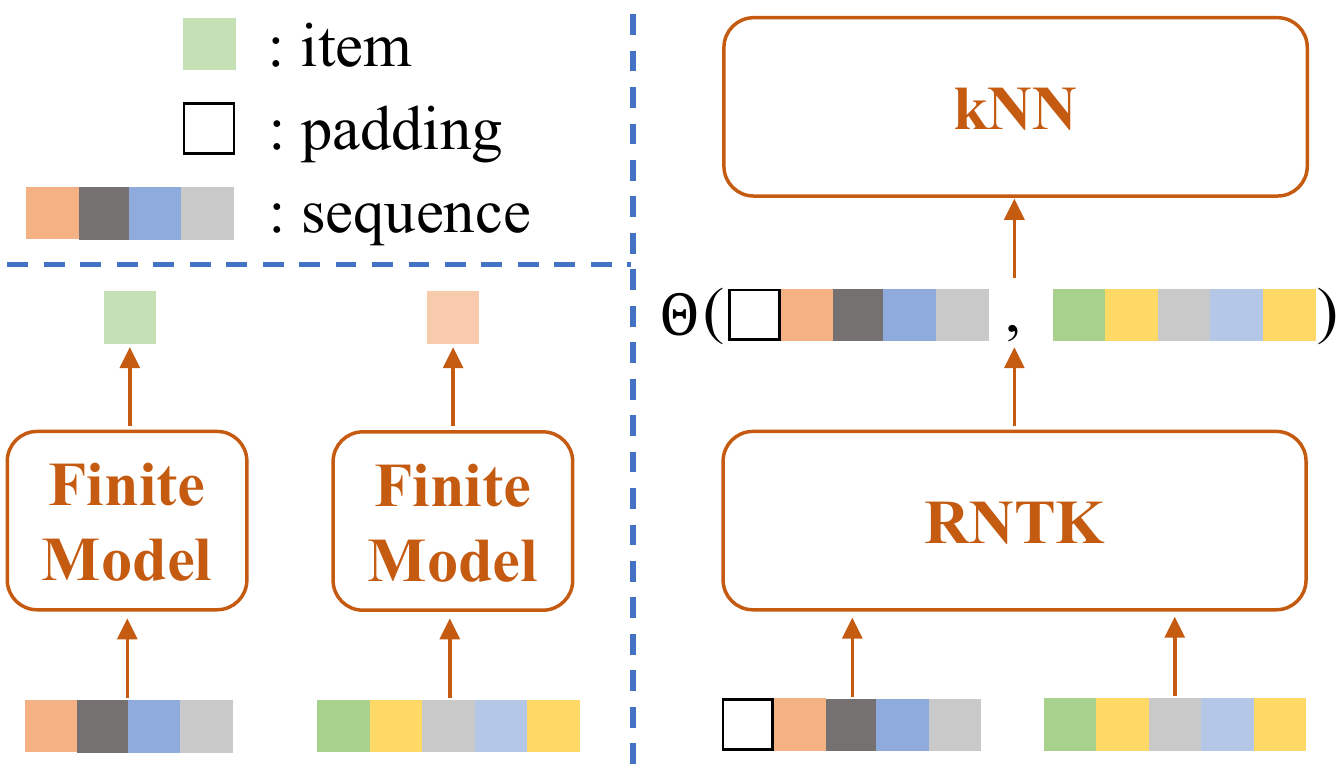}
    \caption{The comparison of general finite models and OverRec. On the left, general finite models build a finite encoder for sequence. On the right, OverRec uses RNTK, $\Theta$, to measure the similarity between sequences, followed by a $k$NN classifier.}
    \label{fig:whole}
\end{figure}

\section{OverRec}
The overview of OverRec is presented in Fig.~\ref{fig:whole}. OverRec utilises RNTK as a similarity measurement, followed by a $k$NN. The whole procedure does not involve any training.

\subsection{Problem Definition}
In sequential recommendations setting, an item set $\mathcal{X}$ contains all items, and $| \mathcal{X}|$ is the number of items. Interactions of a user are represented by a sequence $s=\left[x_1,x_2,\ldots,x_t \right]$, where $x_i\in\mathcal{X},0\leq i\leq t$ and $t$ is the current time step. The sequential recommendation task is to predict next interaction at time step $t+1$, i.e., $x_{t+1}$ for the user. In the following, $\boldsymbol{x}$ is overloaded as the one-hot encoding of $x$.

\subsection{RNTK with Zero Padding}
\label{sec:padding}
In OverRec, the similarity between sequences are calculated by RNTK. From Equation (\ref{eq:nngp-rnn}) and (\ref{eq:rntk}), the analytic calculation is applicable to sequences of different lengths. For efficient batch computing, zero padding is normally applied in the acceleration of sequential problems. In this paper, zero padding is used so that Algorithm~\ref{alg:nngp-rnn} and~\ref{alg:rntk} for sequences with the same length can be applied to user sequences with different lengths.

The most important thing in zero padding for RNTK is that directly padding zero will result in incorrect kernels. Consider two independent one-hot embeddings, whose inner product is zero. In Algorithm~\ref{alg:nngp-rnn}, although the inner product term is zero in Line~\ref{alg:nngp-rnn-dot1} and~\ref{alg:nngp-rnn-dot2}, there is a $\sigma_b$ term, which will contribute to the result. In Algorithm~\ref{alg:rntk}, the derivative term from $V_{\phi}'$ will also have a nonzero result. If adding padding to the sequences and directly fitting them into these two algorithms, the result will include the interactions of paddings and the elements in the original sequence, carrying nonzero terms such as $\sigma_b$ and $V_{\phi}'$. This is unexpected nonzero terms that will affect the kernels.

To benefit from the acceleration and avoid the unwanted padding issue, the following measures are required: (1) The initial hidden state $\boldsymbol{h}^{(l,0)}$ is set to zero. This setting can avoid the variance $\sigma_h$ being carried through zero paddings. (2) The variance $\sigma_b$ of parameters $\boldsymbol{b}$ in Equation (\ref{eq:init}) is set to zero. This setting prevents the variance from aggregating in NNGP-RNN. (3) In the activation function calculation, when there is a term in $\boldsymbol{K}$ is zero padding, both $V_\phi$ and $V_{\phi'}$ will be set as zero. This is because when there is a term of padding, there should not be any forward or backward calculation originally. All of these settings are general and common in existing sequential recommendation models with finite neural network~\cite{gru4rec,sasrec}.

\subsection{$k$-Nearest Neighbours}
With the above developed RNTK as similarity measurement for sequences, a $k$-nearest neighbours ($k$NN) is applied to perform the prediction for the next item.

Given an input sequence $\boldsymbol{x}$, the most similar $k$ sequences of $\boldsymbol{x}$ can be selected based on the RNTK. Using weighted $k$NN, the prediction $\hat{\boldsymbol{y}}$ of $\boldsymbol{x}$ can be defined as:
\begin{equation}
\label{eq:weighted_y}
    \hat{\boldsymbol{y}}=\frac{\sum_{i=1}^k\Theta(\boldsymbol{x},\boldsymbol{x}_i)\boldsymbol{y}_i}{\sum_{i=1}^k\Theta(\boldsymbol{x},\boldsymbol{x}_i)},
\end{equation}
where $\boldsymbol{y}\in\mathbb{R}^{|\mathcal{X}|}$ is the target of neighbouring sequences.

\section{Discussion}
In this section, the following aspects will be discussed: (1) relationship between OverRec and infinite RNN models; (2) relationship between OverRec and other $k$NN methods.

\subsection{Relationship with Infinite RNN}
\label{sec:rel-inrnn}
Although OverRec is based on the RNTK, there are some differences between OverRec and infinite RNN, and the ability of OverRec is restricted by the modelling capability of RNN.

First, OverRec inherits the encoding ability of infinite RNN but not directly the prediction result. According to Equation (\ref{eq:lin}), the derivative of the parameter serves as a feature encoding function for the approximated linear model for a neural network. To obtain a prediction based on this encoding, an extra linear model is required. From the perspective of traditional machine learning, this extra feature encoding module can be integrated into kernel-based methods such as kernel Support-vector Machine (SVM)~\cite{svm}. This extra training with SVM using the NTK kernels is commonly seen in existing methods~\cite{ntk-exact,gntk,rntkuci}. However, because of the well-know difficulty to scale up a kernel SVM (generally applicable to $10^3\sim10^4$ data points), it is impractical to apply kernel SVM for sequential recommendation (usually $10^5\sim10^6$ sequences for open source datasets). A further obstacle is that in sequential recommendation, the prediction result requires to output a ranking over the whole item set. This is extremely time and resource exhausted to calculate for a kernel SVM to output every probability for each prediction class. Therefore, a $k$NN-based prediction is applied in OverRec, which will be inferior in prediction compared with a kernel SVM or the equivalent infinite RNN due to the lack of a learnable prediction layer.

Another difference between OverRec and infinite/large-width RNN in sequential recommendation is that the initialisation of the parameters is different. According to Equation (\ref{eq:init}), there is a $1/\sqrt{n}$ term in initialisation, which is rarely used in existing sequential recommendation methods. This setting is crucial for the convergence of NTK and RNTK. Under this situation, the encoding property and the  recommendation performance will be affected by the difference in initialisation as well.

\subsection{Relationship with $k$NN Methods}
\label{sec:rel-knn}
OverRec applies a $k$NN for prediction, which utilises a similar prediction procedure as S$k$NN~\cite{sknn} and STAN~\cite{stan}. The major differences are the similarity measurement of sequences and the way to use the similarity to perform recommendation.

In terms of the similarity measurement of sequences, the most important factor is to develop an accurate sequential measurement for semantic similarity. In defining the semantic similarity, OverRec is originated from infinite RNN, which means that the semantic similarity in OverRec is equivalent to the similarity metric in the latent space of RNN output. While for S$k$NN and STAN, the similarity is based on counting the co-occurrence of items in different sequences. This difference in defining the semantic similarity is mainly accountable for the different behaviours for all of these $k$NN-based methods.

As for the calculation of recommendation ranking, OverRec directly generates the prediction based on the target items of semantically similar sequences. For S$k$NN and STAN, the ranking of items is considering all the items in the similar sequences. The difference in choosing the target item reflects the accuracy of measuring the semantic similarity. For a more accurate semantic similarity based on infinite RNN, OverRec can directly rely on the target items from similar sequences. While for S$k$NN and STAN, since the sequence similarity is based on item co-occurrence, which is not accurate enough, solely using the target items from these sequences are not reliable for generating recommendation ranking.

\section{Experiment}
\label{sec:exp}
In experiments, we answer these research questions (RQ):
\begin{itemize}
    \item \textbf{RQ1}: How does OverRec perform compared with the state-of-the-art methods? (Section~\ref{sec:rq1})
    \item \textbf{RQ2}: How does OverRec perform compared with finite RNN models? (Section~\ref{sec:rq-rnn})
    \item \textbf{RQ3}: How does OverRec perform compared with other $k$NN methods? (Section~\ref{sec:rq-knn})
    \item \textbf{RQ4}: How is the sensitivity of the hyper-parameters in OverRec model? (Section~\ref{sec:rq-param})
\end{itemize}

\begin{table}[t]
    \caption{Statistics of the datasets after preprocessing.}
    \centering
    \resizebox{0.8\linewidth}{!}{
    \begin{tabular}{l|rrrrr}
    \toprule
    Specs. & Beauty & Clothing & Toys & ML-1M  \\
    \midrule
    $\sharp$ Users & 22,363 & 39,387 & 19,413 & 6,041\\
    $\sharp$ Items & 12,101 & 23,033 & 11,925 & 3,417\\
    $\sharp$ Avg. Length & 8.9 & 7.1 & 8.6 & 165.5\\
    $\sharp$ Actions & 198,502 & 278,677 & 167,597 & 999,611\\
    Sparsity & 99.93\% & 99.97\% & 99.93\% & 95.16\%\\
    \bottomrule
    \end{tabular}
    }
    \label{tab:datasets}
\end{table}

\begin{table*}[t]
    \caption{Overall performance. Bold scores represent the highest results of all methods. Underlined scores stand for the highest results from previous methods. Since OverRec is a deterministic algorithm, the standard deviation of the results is inapplicable. Both OverRec and the NNGP-RNN are our proposed methods.}
    \centering
    \resizebox{0.9\linewidth}{!}{
    \begin{tabular}{c|l|ccccccc|c|c|c}
    \toprule
         Dataset& Metric & BPR-MF & Item-$k$NN & GRU4Rec & SASRec & BERT4Rec &S$k$NN &STAN&NNGP-RNN& OverRec&Improvement\\
         \midrule
         \multirow{4}*{Beauty}&MRR@5&0.0024&0.0018&0.0063&\underline{0.0194}&0.0153&0.0035&0.0083&0.0203&\textbf{0.0217}&+11.86\%\\
         &MRR@10&0.0037&0.0032&0.0089&\underline{0.0213}&0.0172&0.0071&0.0107&0.0220&\textbf{0.0238}&+11.74\%\\
         &NDCG@5&0.0040&0.0037&0.0086&\underline{0.0236}&0.0187&0.0057&0.0126&0.0254&\textbf{0.0263}&+11.44\%\\
         &NDCG@10&0.0053&0.0051&0.0142&\underline{0.0281}&0.0268&0.0145&0.0185&0.0304&\textbf{0.0312}&+11.03\%\\
         \midrule
         \multirow{4}*{Clothing}&MRR@5&0.0027&0.0023&0.0045&\underline{0.0068}&0.0052&0.0015&0.0024&0.0070&\textbf{0.0072}&+5.88\%\\
         &MRR@10&0.0041&0.0034&0.0054&\underline{0.0080}&0.0068&0.0037&0.0032&0.0078&\textbf{0.0081}&+1.25\%\\
         &NDCG@5&0.0052&0.0048&0.0061&\underline{0.0091}&0.0075&0.0026&0.0037&0.0086&\textbf{0.0089}&-2.20\%\\
         &NDCG@10&0.0069&0.0062&0.0083&\underline{0.0124}&0.0102&0.0080&0.0057&0.0106&\textbf{0.0111}&-10.48\%\\
         \midrule
        \multirow{4}*{Toys}&MRR@5&0.0025&0.0021&0.0067&\underline{0.0248}&0.0232&0.0047&0.0092&0.0265&\textbf{0.0272}&+9.68\%\\
         &MRR@10&0.0036&0.0033&0.0079&\underline{0.0283}&0.0276&0.0089&0.0120&0.0286&\textbf{0.0292}&+3.18\%\\
         &NDCG@5&0.0053&0.0046&0.0059&\underline{0.0306}&0.0293&0.0079&0.0142&0.0315&\textbf{0.0323}&+5.56\%\\
         &NDCG@10&0.0067&0.0058&0.0084&\underline{0.0374}&0.0358&0.0181&0.0212&0.0366&\textbf{0.0376}&+0.53\%\\
         \midrule
         \multirow{4}*{ML-1M}&MRR@5&0.0078&0.0053&0.0269&\underline{0.0514}&0.0326&0.0024&0.0162&0.0573&\textbf{0.0585}&+13.81\%\\
         &MRR@10&0.0162&0.0097&0.0313&\underline{0.0594}&0.0382&0.0046&0.0249&0.0643&\textbf{0.0650}&+9.43\%\\
         &NDCG@5&0.0052&0.0046&0.0385&\underline{0.0638}&0.0432&0.0039&0.0258&0.0699&\textbf{0.0711}&+11.44\%\\
         &NDCG@10&0.0079&0.0069&0.0671&\underline{0.0910}&0.0619&0.0094&0.0466&0.0863&\textbf{0.0866}&-4.84\%\\
         \bottomrule
    \end{tabular}
    }
    \label{tab:overall}
\end{table*}

\subsection{Setup}
\label{sec:setup}
\subsubsection{Dataset}
The experiments are conducted over four benchmark datasets as shown in Table~\ref{tab:datasets}.
\begin{itemize}
    \item \textbf{Amazon Beauty}, \textbf{Clothing} and \textbf{Toys}~\cite{amazon}\footnote{http://jmcauley.ucsd.edu/data/amazon/}. Following baselines~\cite{sasrec,bert4rec,s3rec}, the widely used Amazon dataset is chosen in our experiments with three sub-categories.
    \item \textbf{MovieLens-1M}~\cite{movielens}\footnote{https://grouplens.org/datasets/movielens/1m/}. Following~\cite{bert4rec}, the popular movie recommendation dataset is used here, denoted as ML-1M.
\end{itemize}

Following~\cite{sasrec,bert4rec,s3rec}, all interactions are considered as implicit feedback. Users or items appearing less than five times are removed. The maximum length of a sequence is 50.

\subsubsection{Metrics}
\label{sec:metric}
To evaluate the performance of different models, top-$k$ Mean Reciprocal Rank (MRR@$k$) and top-$k$ Normalised Discounted Cumulative Gain (NDCG@$k$) are used, where $k$ are chosen from $\{5,10\}$. The evaluation of the ranking results are conducted over the entire item set for the fair comparison~\cite{metric}.

\subsubsection{Baselines}
\label{sec:baseline}
The following methods are compared:
\begin{itemize}
    \item \textbf{BPR-MF}~\cite{bprmf} is a method trains a matrix factorisation model using BPR.
    \item \textbf{Item-$k$NN}~\cite{item-knn} is a method based on item nearest neighbours. It does not consider sequential information.
    \item \textbf{GRU4Rec}~\cite{gru4rec} applies GRU as a recurrent model to encode user sequences.
    \item \textbf{SASRec}~\cite{sasrec} is an attention-based model in single direction.
    \item \textbf{BERT4Rec}~\cite{bert4rec} uses a masked item training scheme to train a bi-directional self-attention model.
    \item \textbf{S$k$NN}~\cite{sknn} is a method calculating the cosine similarity of different sequences and consider all the items in the neighbouring sequences.
    \item \textbf{STAN}~\cite{stan} is an extension of S$k$NN to incorporate time effect in similarity and prediction calculations.
\end{itemize}

\subsubsection{Implementation}
\label{sec:imple}
To make batch acceleration possible, $\sigma_b$ is set to zero. For the rest $\sigma_{W},\sigma_{U}$ and $\sigma_{V}$, they are chosen from $\{0.3,0.5,0.8,1,1.2,\sqrt{2}\}$. The $k$ for $k$NN is chosen according to the top-$k$ evaluation ranking metric.

\subsection{Overall Performance}
\label{sec:rq1}
Performances of OverRec and baselines are in Table~\ref{tab:overall}.

Overall, although OverRec is method without any training, it outperforms the state-of-the-art methods such as SASRec and BERT4Rec. When comparing OverRec with the finite recurrent model, GRU4Rec, OverRec consistently has a higher ranking performance. This verifies that OverRec can benefit from the \textit{double ascent} phenomenon of infinite-width neural network. The improvements from infinite RNN can even booster OverRec to have a stronger result compared with the more powerful attention model such as SASRec and BERT4Rec. In addition, NNGP-RNN from Equation (\ref{eq:nngp-rnn}) is applied to substitute RNTK in OverRec, which has a slightly lower result. Although NNGP-RNN only has the similar ability to the infinite RNN at initialisation stage, it can still provide a reasonable similarity measurement, mimicking a random feature model.

For traditional non-sequential methods such as BPR-MF and Item-$k$NN, these methods do not work well in sequential recommendation because of the neglect of sequential patterns. The performance is inferior to the neural network-based methods and the sequence-based $k$NN methods.

When RNN is applied in sequential recommendation by GRU4Rec, the performance showed a great improvement due to the more powerful sequential modelling ability. More recently, attention has been introduced in sequential recommendation because it is currently the strongest sequence encoder. SASRec is the first method to apply a uni-directional attention in sequential recommendations. Compared with the previous deep learning models, SASRec can outperform them by a large margin as the strongest baseline in this experiment. This indicates that the model performance is inherently limited by the base sequence encoder. Thus, the performance of OverRec is partially limited by RNN. More recent methods usually use the attention-based encoder as well while training the model with different objectives. For example, BERT4Rec proposes to exploit the masked item prediction objective, which is originated from language models, to enhance the model's ability to infer the contextual semantics by filling in the masks.

In addition, there are two strong sequence similarity-based models, S$k$NN and STAN, which does not require training. Both of these two methods are based on the counting the item co-occurrence in sequences to determine the similarity of sequences. The major difference between S$k$NN and STAN is that STAN assigns different weights to different items according ot the time factor while S$k$NN equally consider the items in sequences. It can be seen from Table~\ref{tab:overall} that S$k$NN can consistently outperform the traditional methods yet there is a gap between S$k$NN and neural network-based methods. For STAN, it is clear that it improves from S$k$NN by a large margin, which indicates that the semantic similarity cannot simply rely on plain co-occurrence of items. Although STAN has improved compared with S$k$NN, its performance is still inferior to neural network-based methods. It can be concluded that the traditional methods and sequential $k$NN-based methods cannot compete with neural network models. However, the proposed OverRec can achieve a comparable result and outperform the existing strong neural models while being a method without training.

\begin{figure}[t]
    \centering
    \subfigure{
    \includegraphics[width=0.27\linewidth]{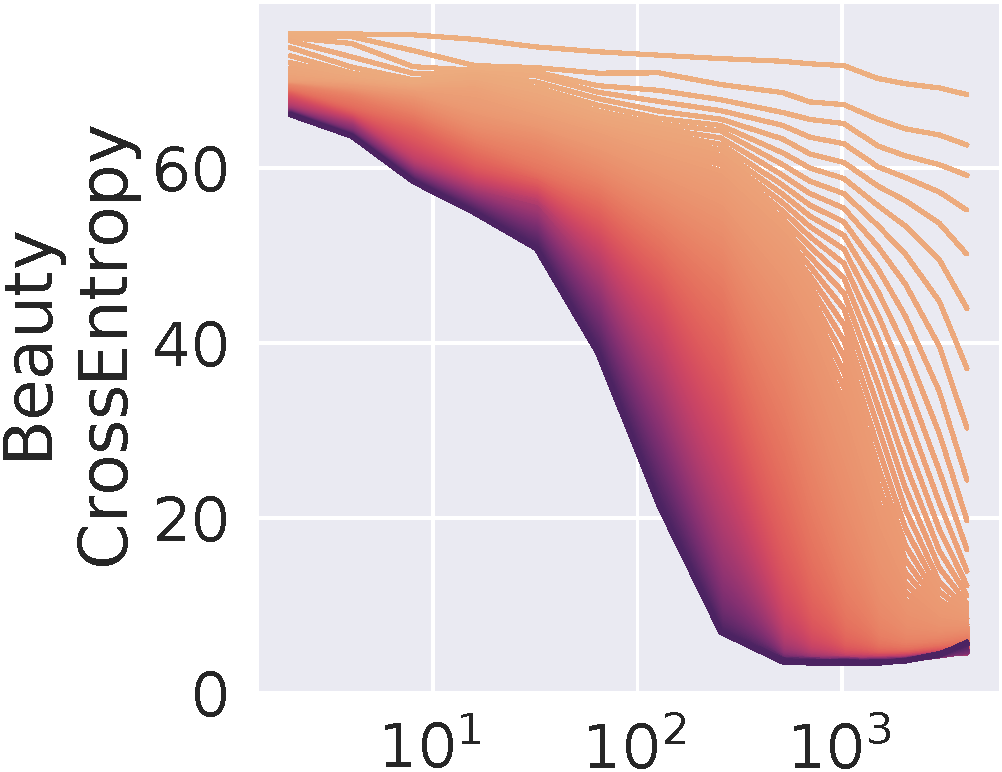}
    }
    \subfigure{
    \includegraphics[width=0.27\linewidth]{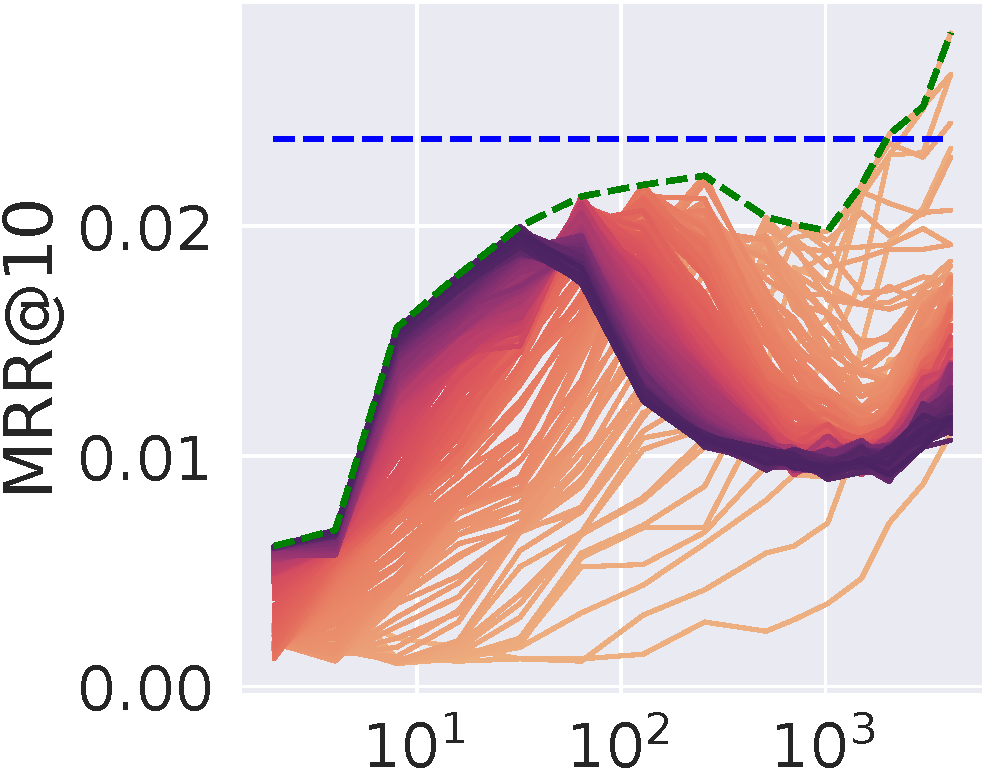}
    }
    \subfigure{
    \includegraphics[width=0.335\linewidth]{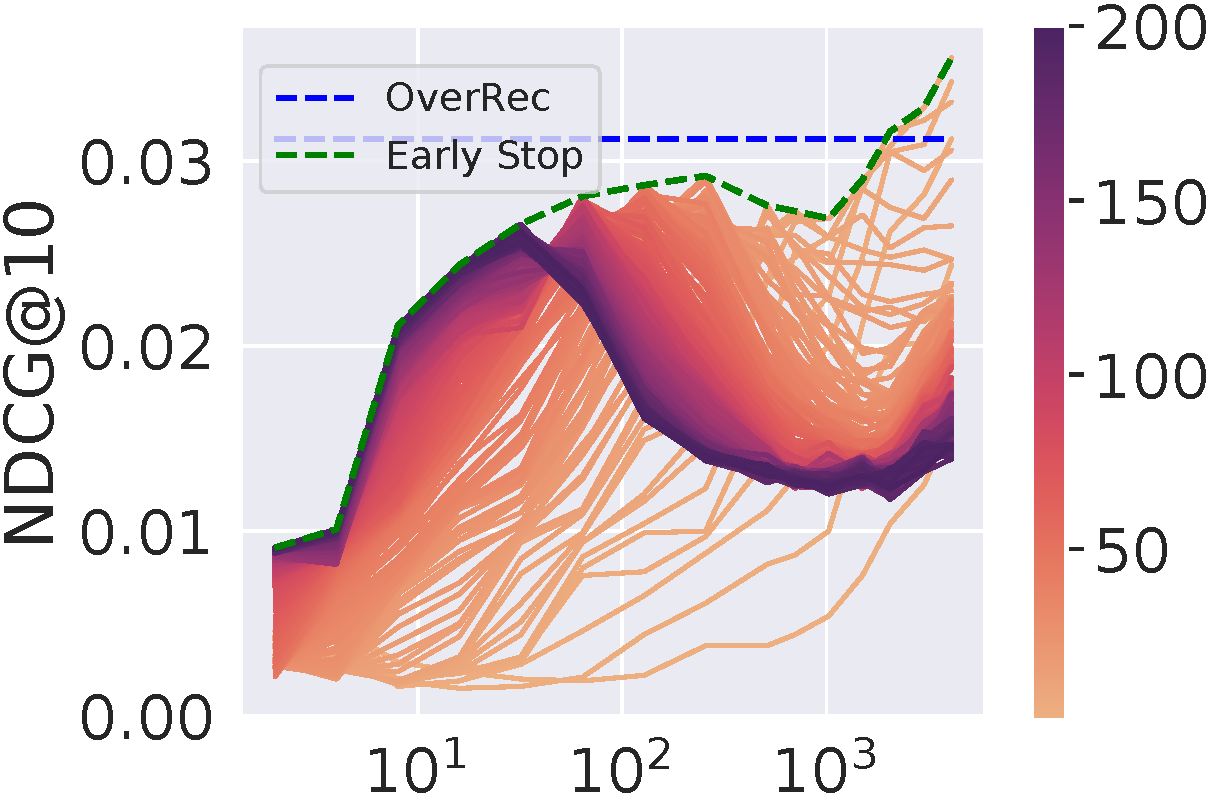}
    }
    \subfigure{
    \includegraphics[width=0.27\linewidth]{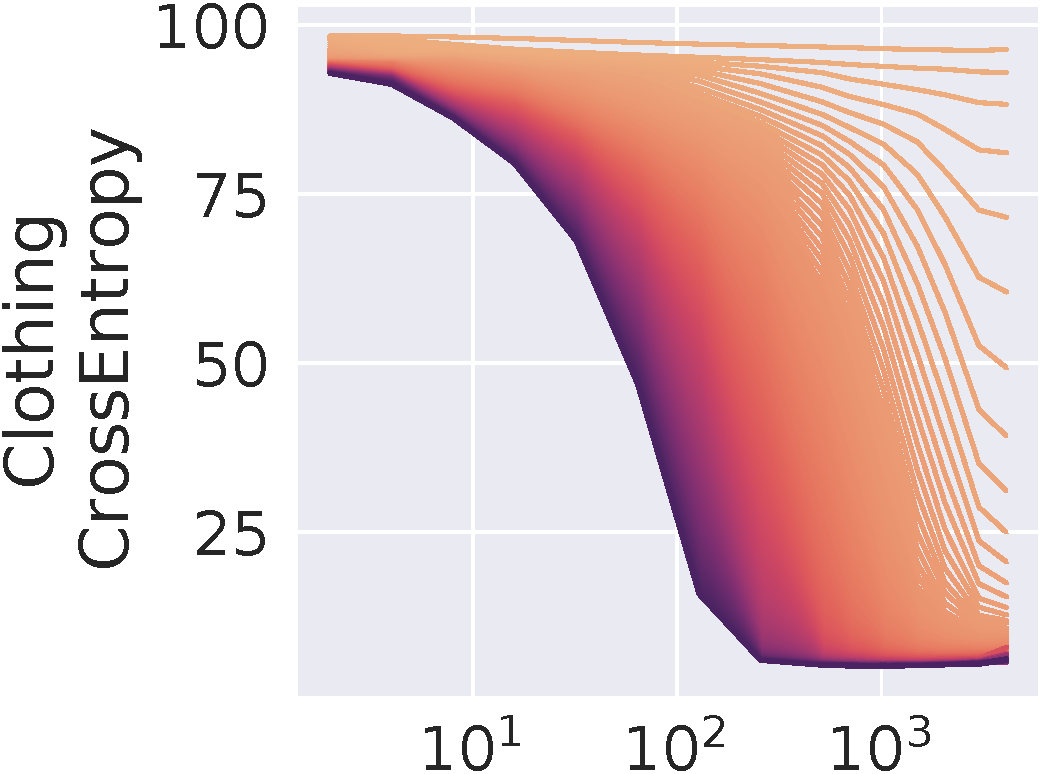}
    }
    \subfigure{
    \includegraphics[width=0.27\linewidth]{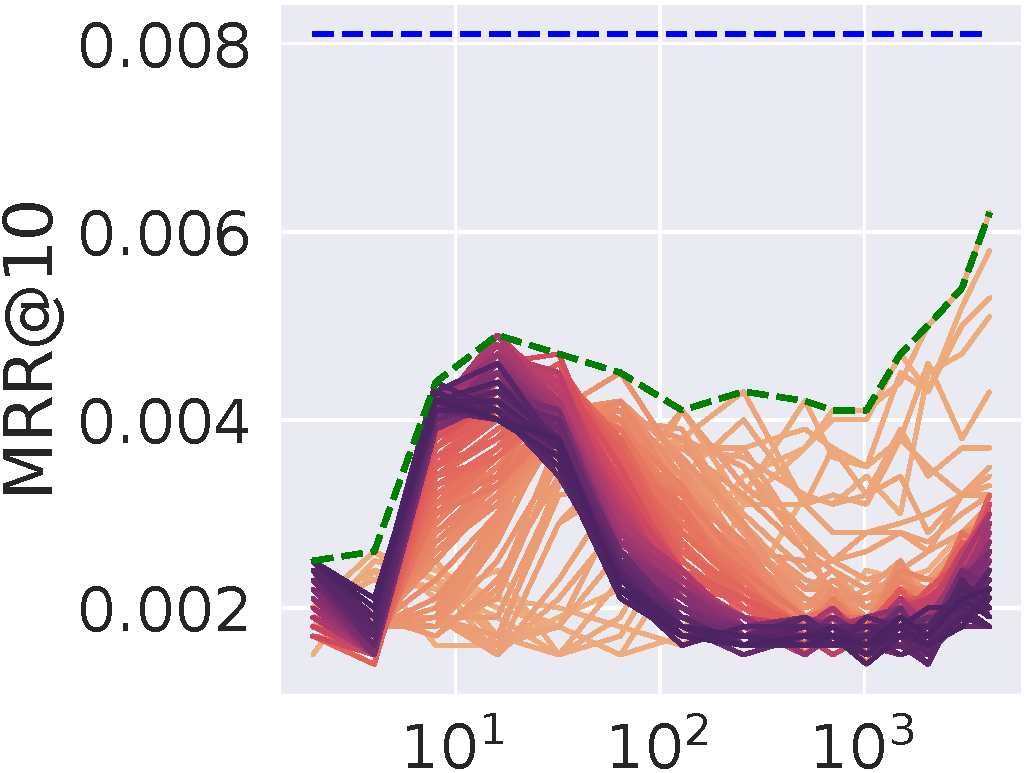}
    }
    \subfigure{
    \includegraphics[width=0.347\linewidth]{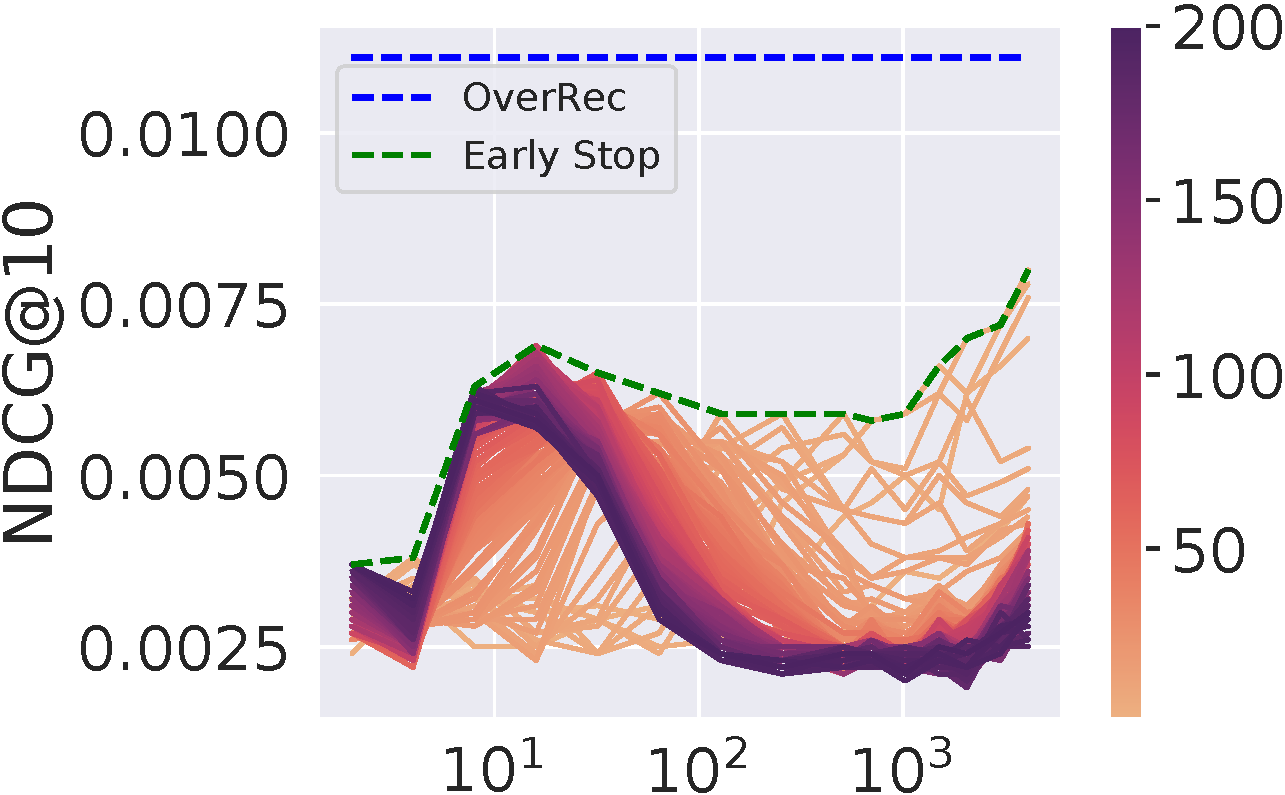}
    }
    \subfigure{
    \includegraphics[width=0.27\linewidth]{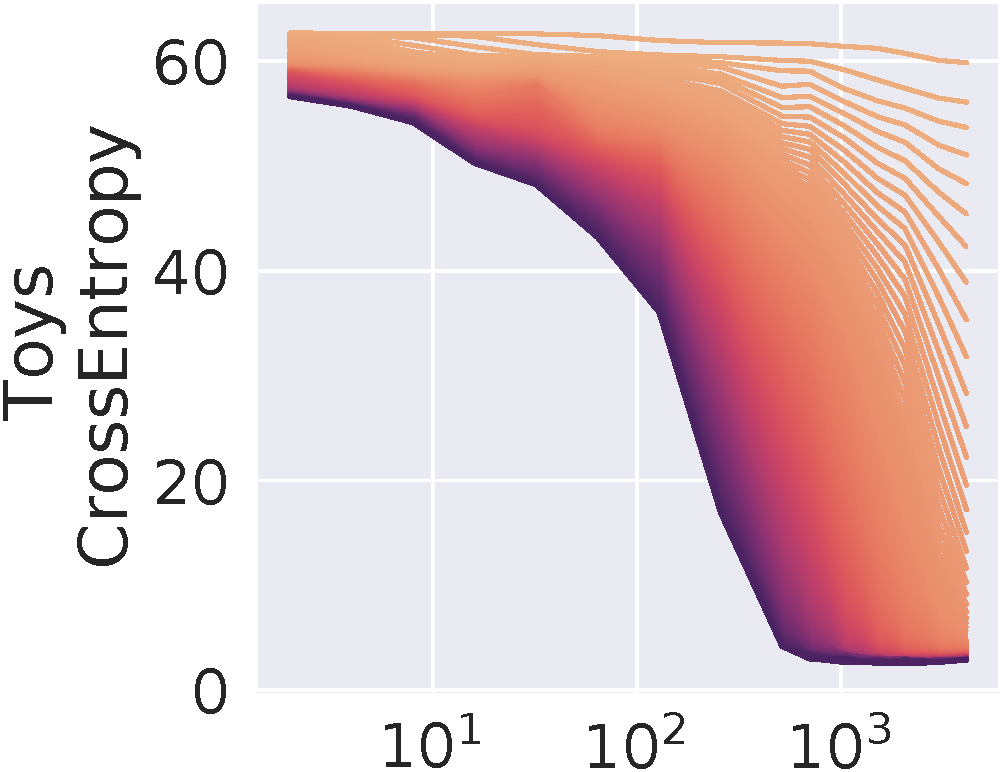}
    }
    \subfigure{
    \includegraphics[width=0.27\linewidth]{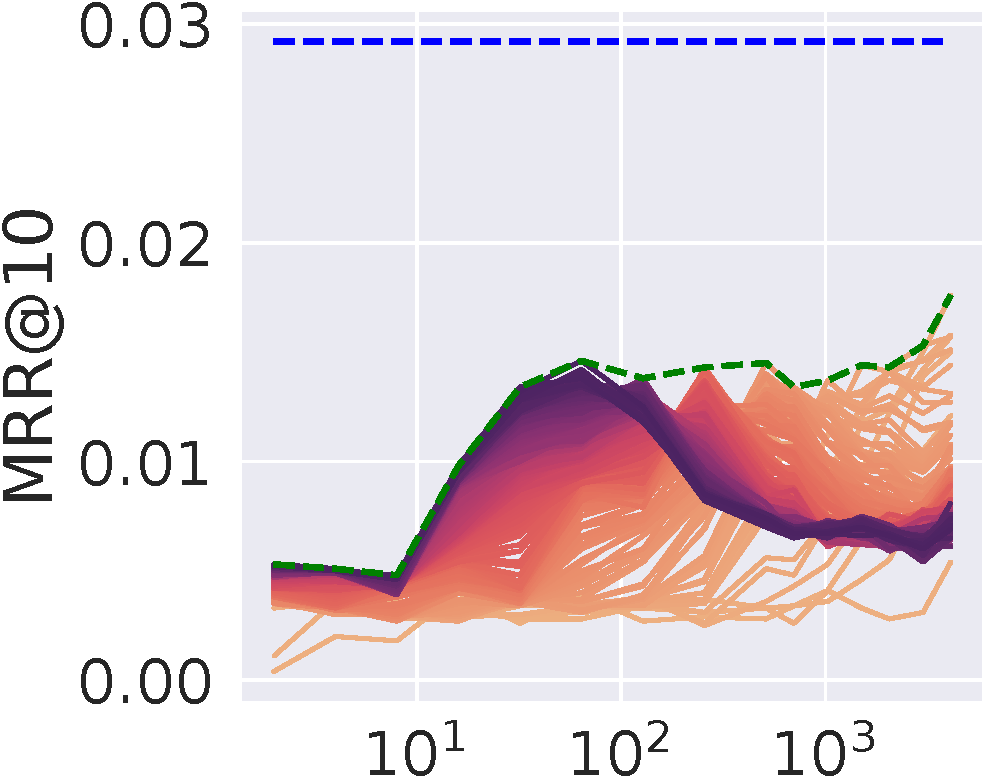}
    }
    \subfigure{
    \includegraphics[width=0.335\linewidth]{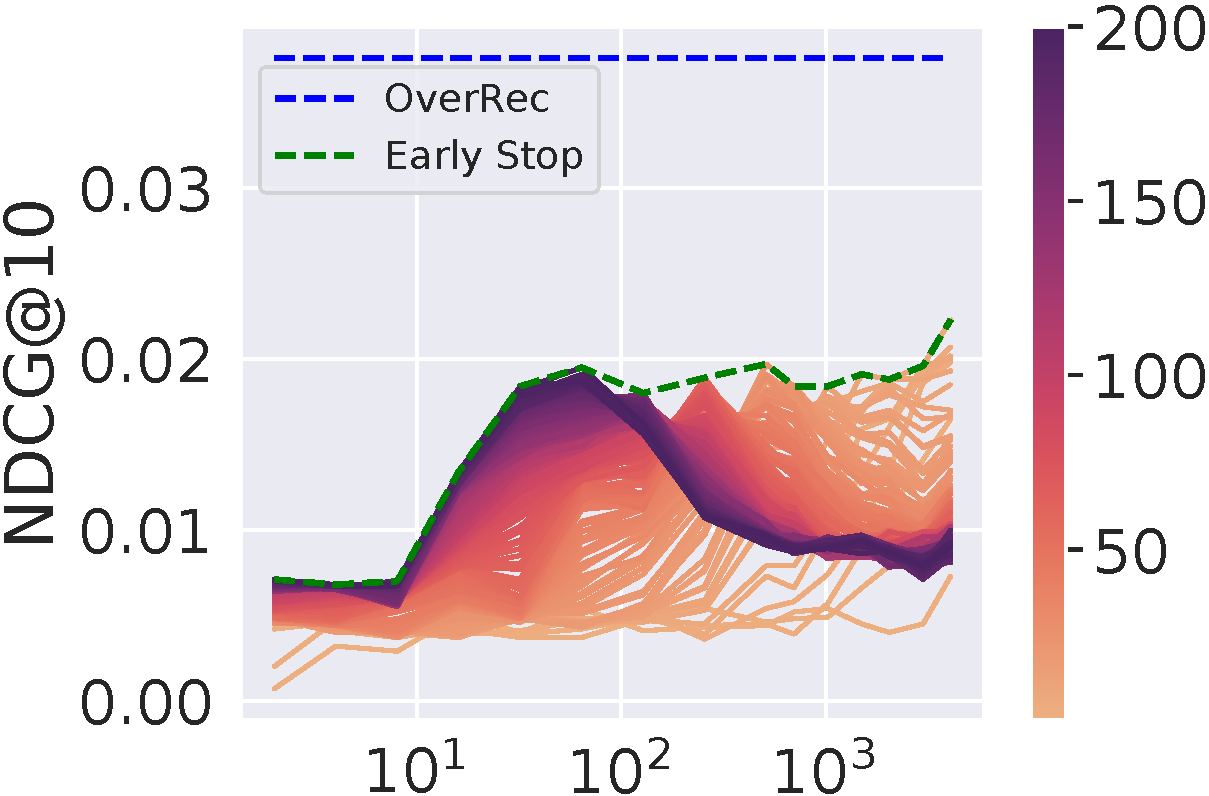}
    }
    \subfigure{
    \includegraphics[width=0.28\linewidth]{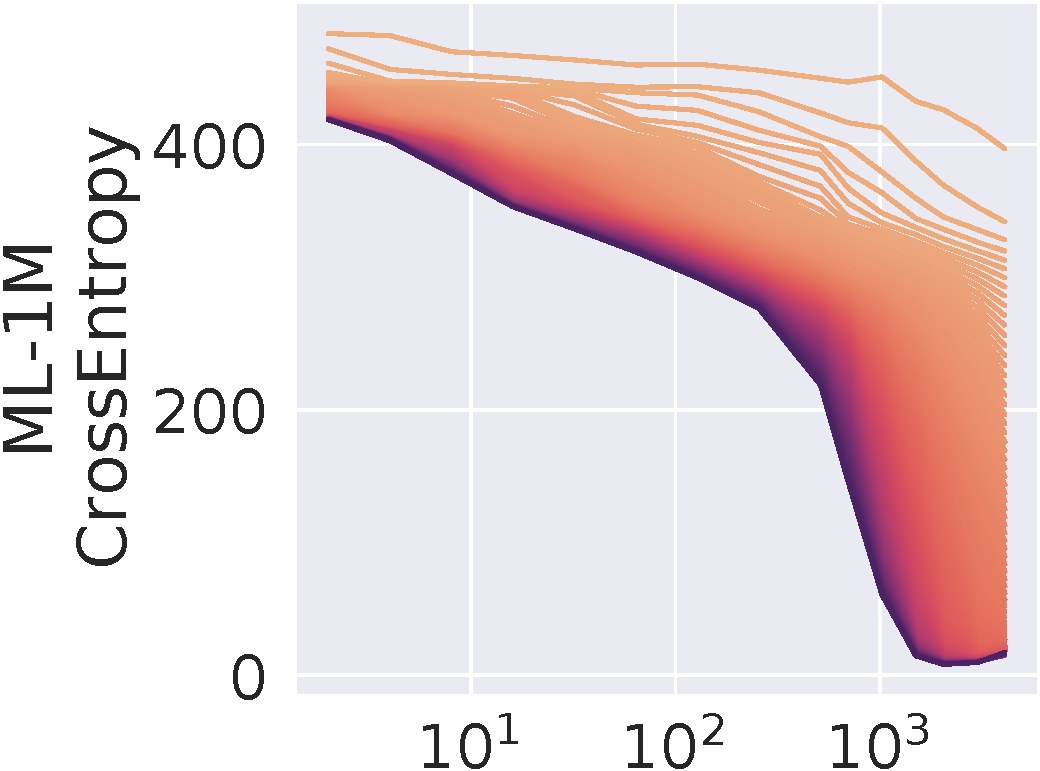}
    }
    \subfigure{
    \includegraphics[width=0.27\linewidth]{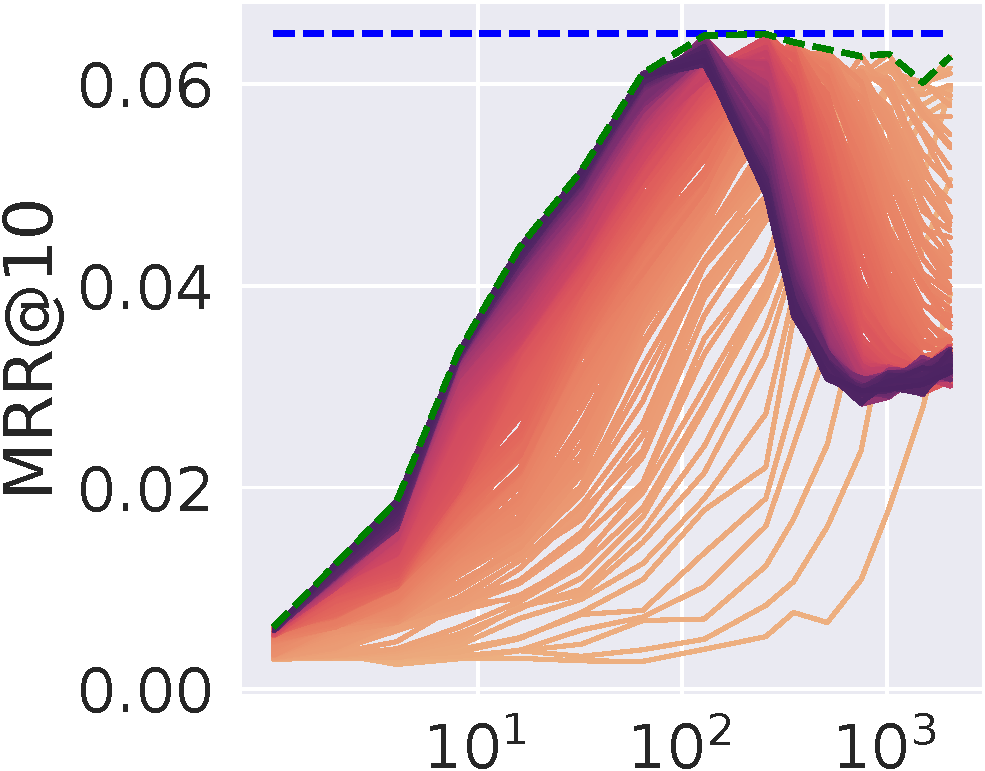}
    }
    \subfigure{
    \includegraphics[width=0.335\linewidth]{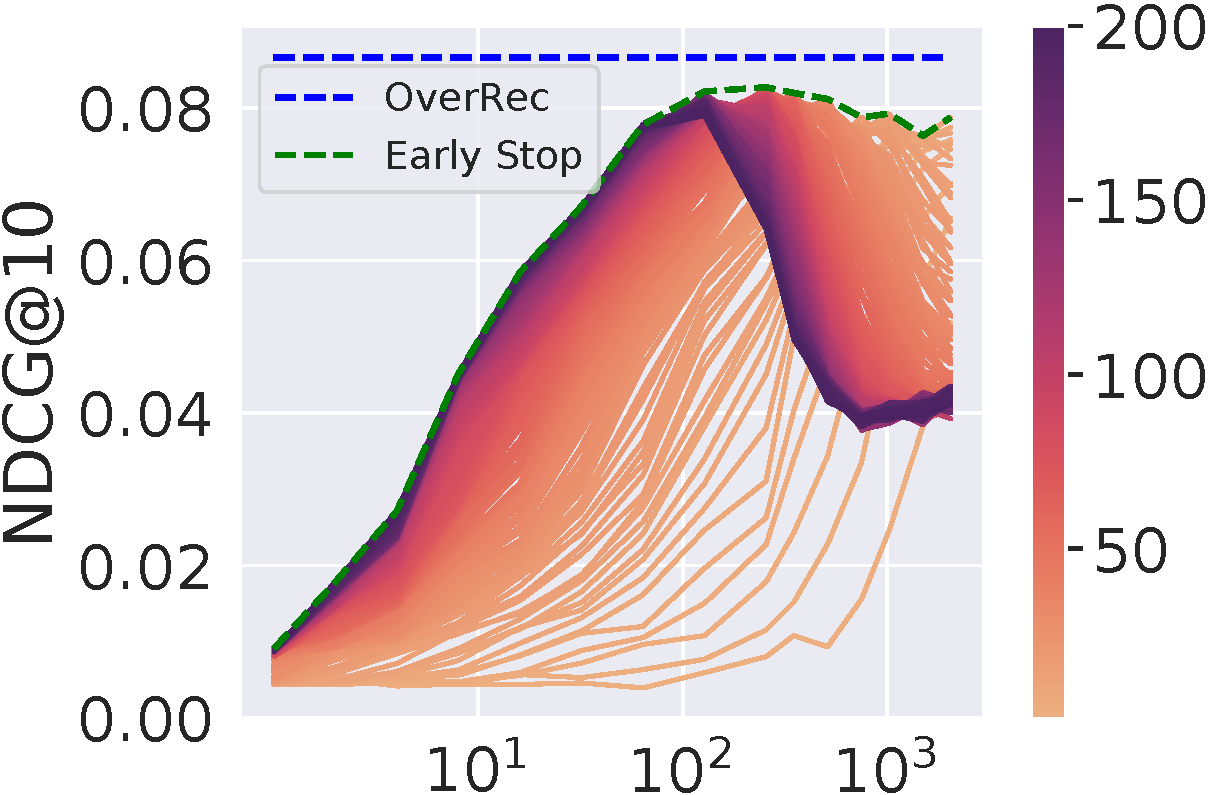}
    }
    \caption{Each row has the training loss and test performances on a dataset. X-axis is the width of models from $2$ to $4096$. The colour bar denotes the training epoch from $1$ to $200$.}
    \label{fig:rnn}
\end{figure}

\subsection{Comparisons with Finite RNN}
\label{sec:rq-rnn}
In this experiment, OverRec is compared with finite RNN with width from $2\sim4096$. Results of training loss, MRR@10 and NDCG@10 are reported in Fig.~\ref{fig:rnn} for all datasets. These results indicate the \textit{double ascent} as well.

The first column demonstrates the training cross entropy loss of RNN models with different widths. From the trend, it is clear that as the width increases, the best training loss of the model becomes lower. For each model, the longer the training is, the lower the training loss it can achieve. This phenomenon indicates that a larger model can achieve a lower training loss compared with smaller models.

In the second and the third columns, the test performance of MRR@10 and NDCG@10 are shown along with the performance of OverRec (blue dashed line) and when applying early stopping for each finite RNN models (green dashed line). The \textit{double ascent} phenomenon is clearly shown in the trend of model performance. No matter for the early stopping performance or the performance at the end of training, the \textit{double ascent} appears in both of these two situations.

Comparing OverRec with finite RNN models, OverRec can perform closely to very large RNN models. Since OverRec is derived from infinite-width RNN, the performance of OverRec is limited to the RNN structure itself as discussed in Section~\ref{sec:rel-inrnn}. Therefore, the close performance aligns with the discussion. While there are occasions that the performance of OverRec is lower than finite RNN as in Beauty dataset, the reason is due to the prediction function in OverRec is based on $k$NN. As discussed in Section~\ref{sec:rel-knn}, $k$NN can lead to a loss of prediction accuracy due to its training scheme.

\begin{figure}[t]
    \centering
    \subfigure[Beauty.]{
    \includegraphics[width=0.35\linewidth]{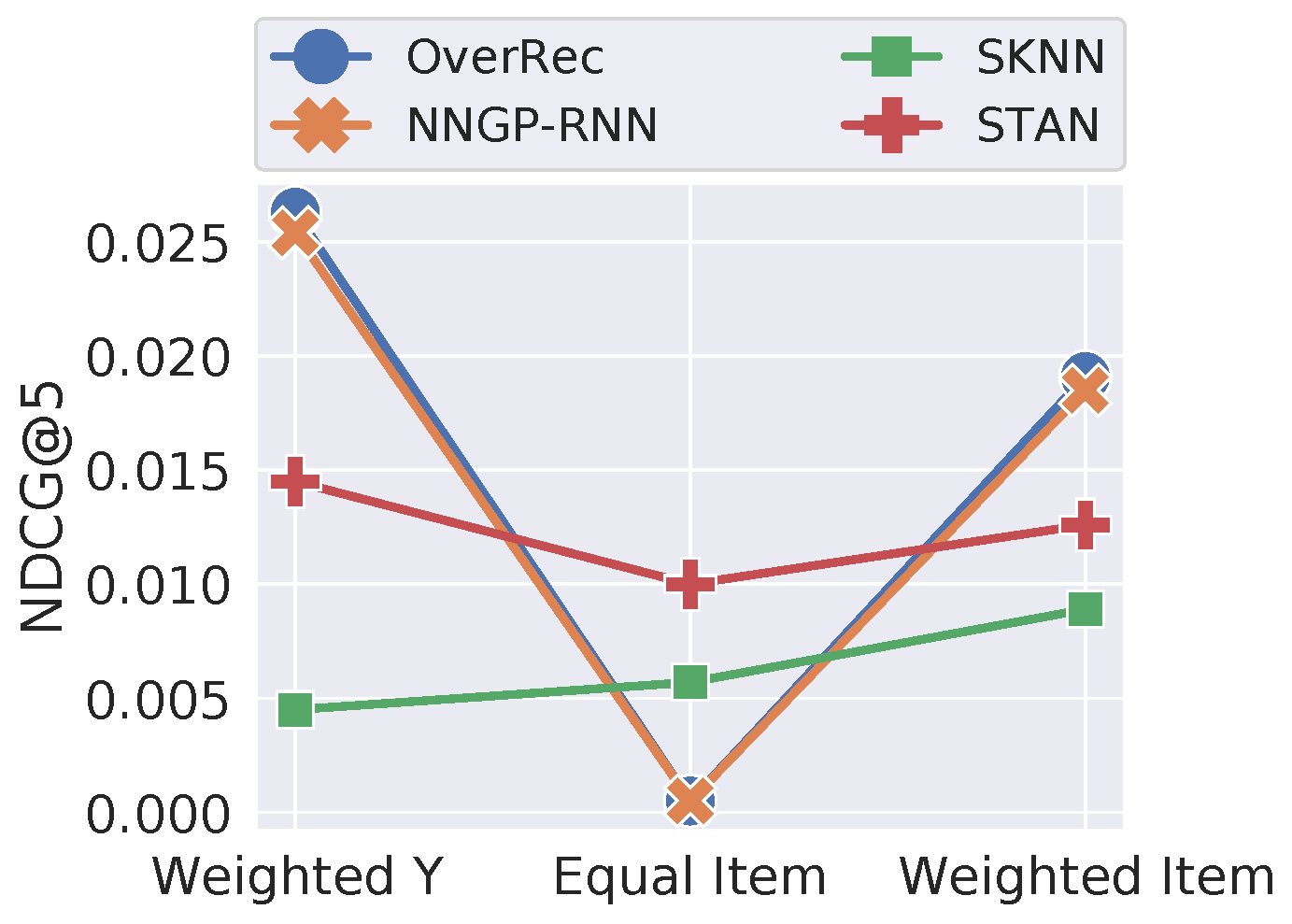}
    }
    \subfigure[Clothing.]{
    \includegraphics[width=0.35\linewidth]{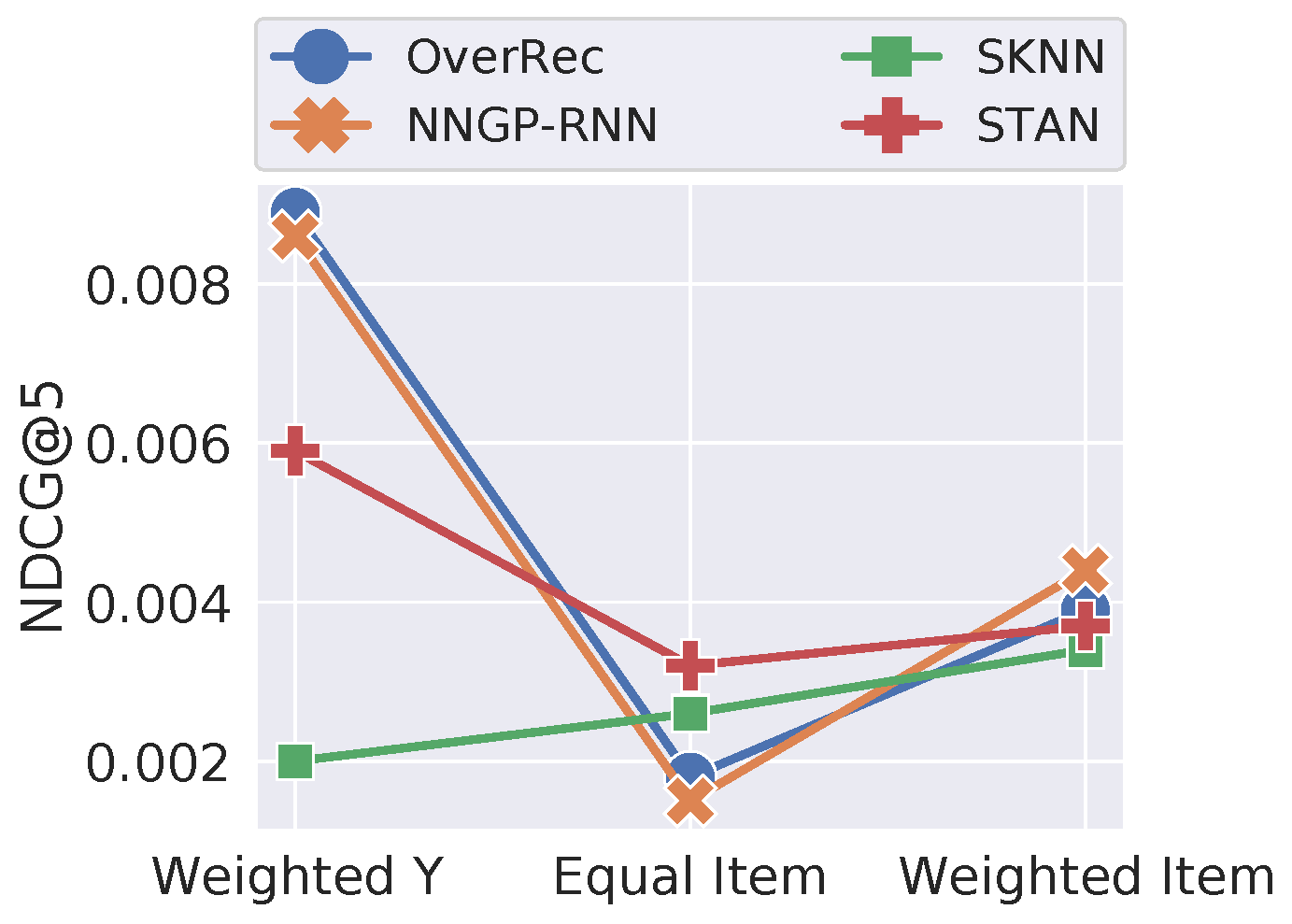}
    }
    \subfigure[Toys.]{
    \includegraphics[width=0.35\linewidth]{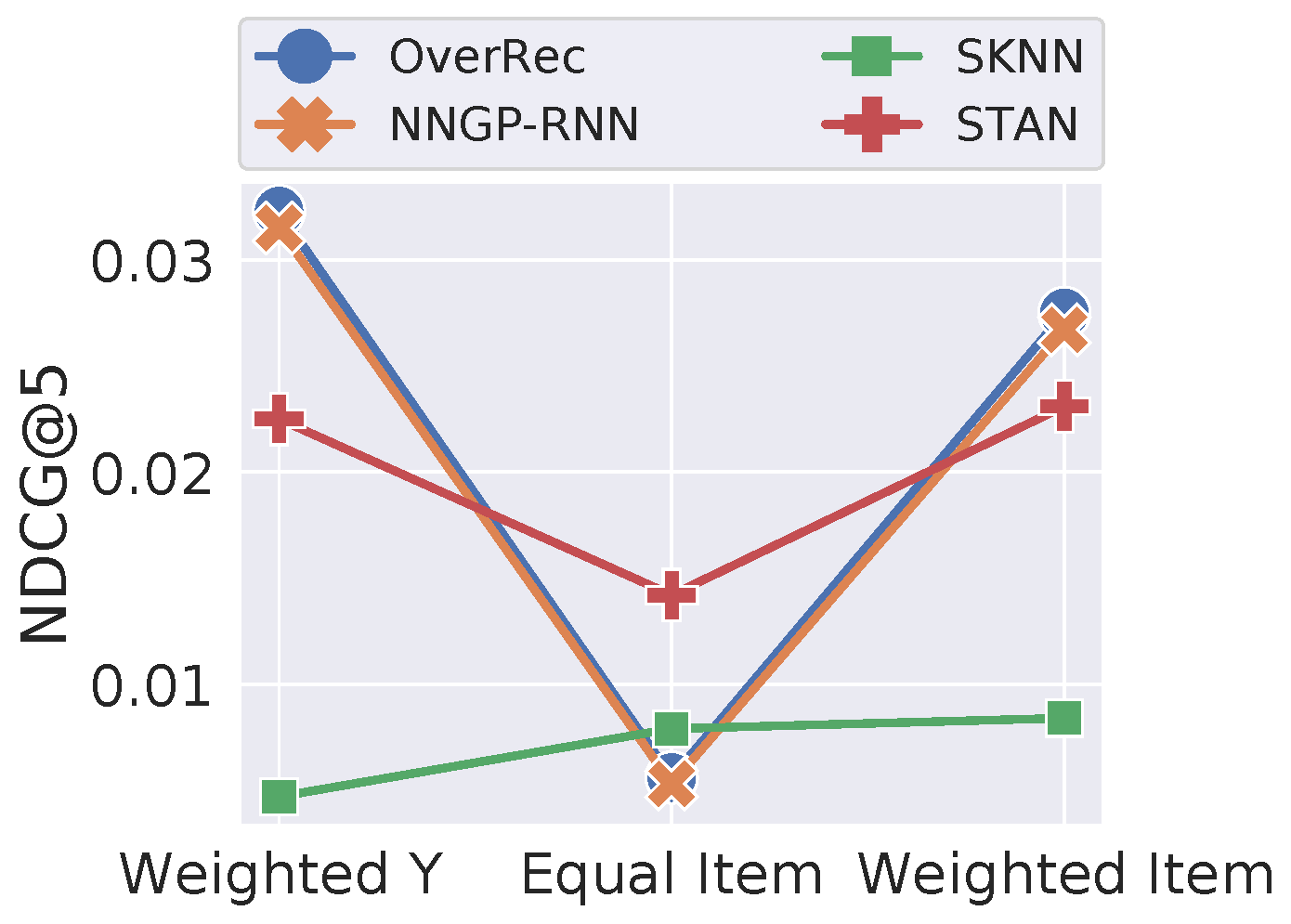}
    }
    \subfigure[ML-1M.]{
    \includegraphics[width=0.35\linewidth]{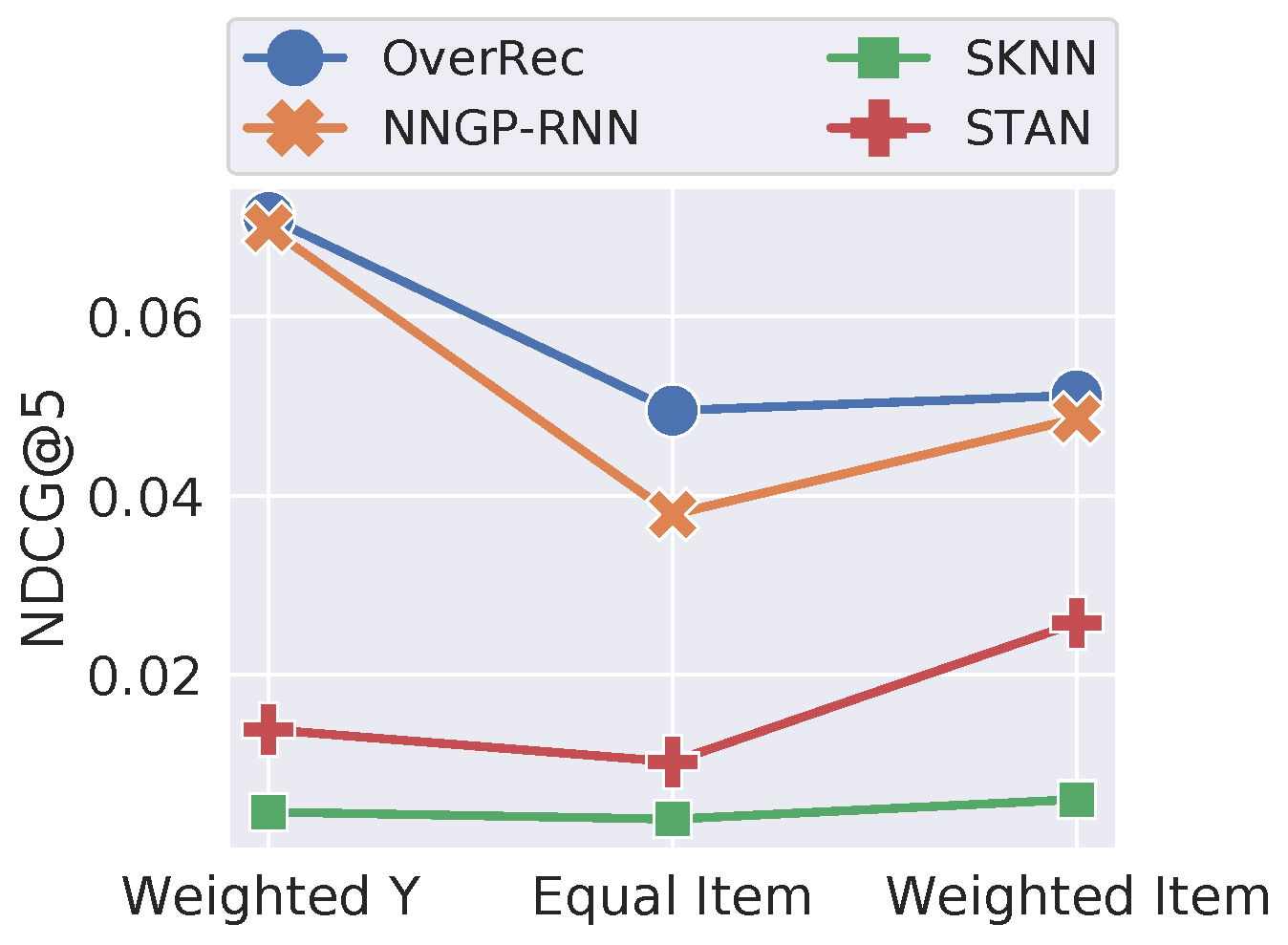}
    }
    \caption{NDCG@5 of different $k$NN methods.}
\label{fig:knn}
\end{figure}

\begin{figure*}[t]
    \centering
    \subfigure[OverRec on Beauty.]{
    \includegraphics[width=0.142\linewidth]{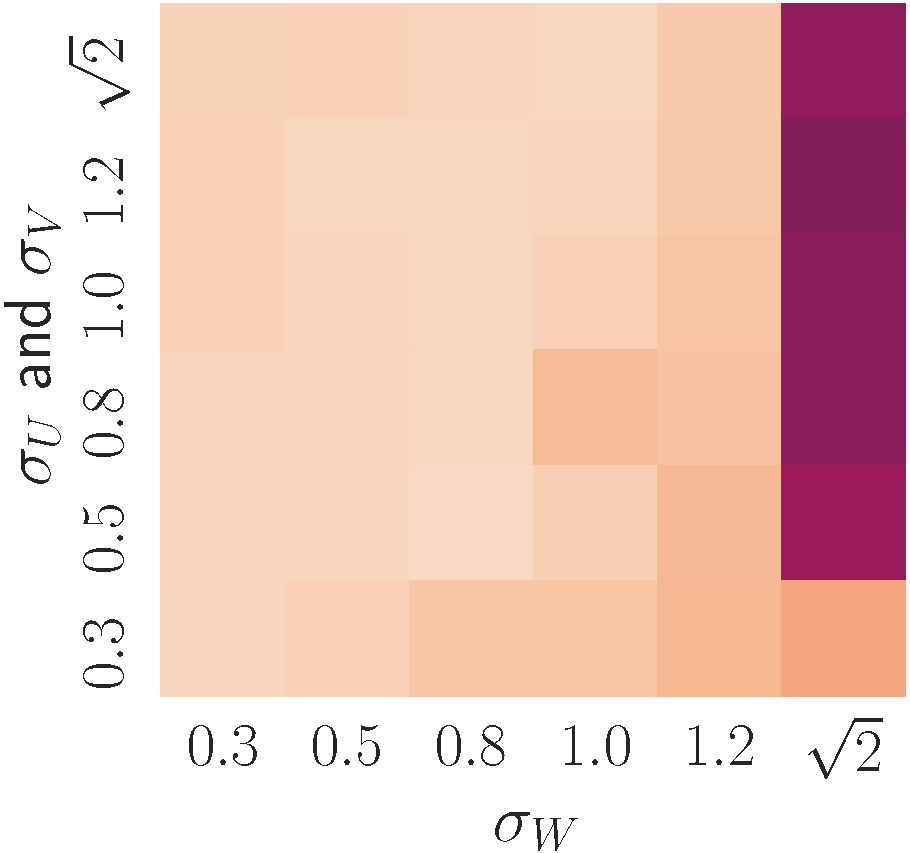}
    }
    \subfigure[NNGP-RNN on Beauty.]{
    \includegraphics[width=0.18\linewidth]{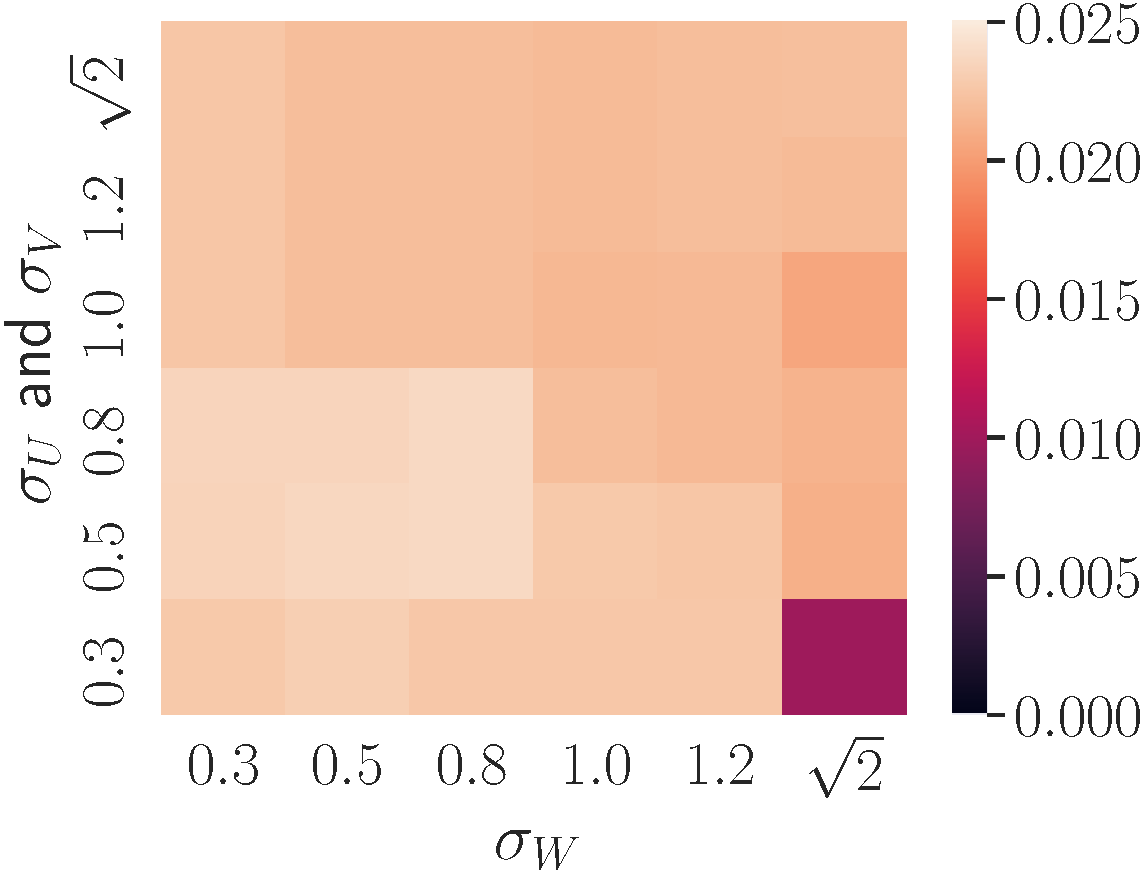}
    }
    \subfigure[OverRec on Clothing.]{
    \includegraphics[width=0.143\linewidth]{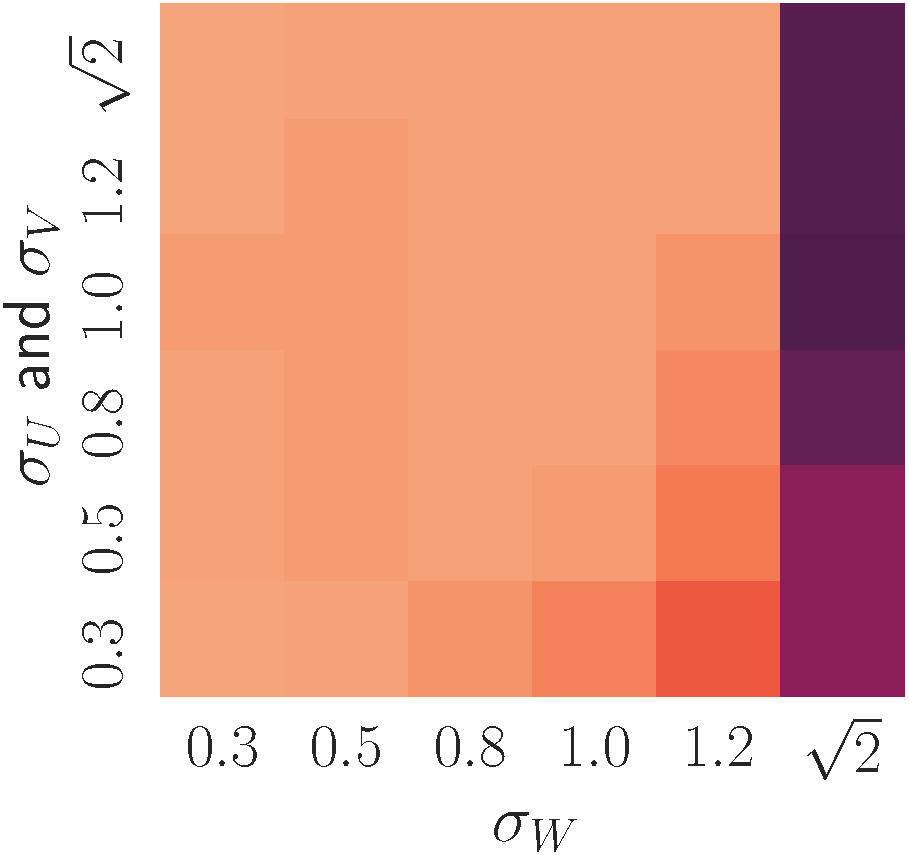}
    }
    \subfigure[NNGP-RNN on Clothing.]{
    \includegraphics[width=0.18\linewidth]{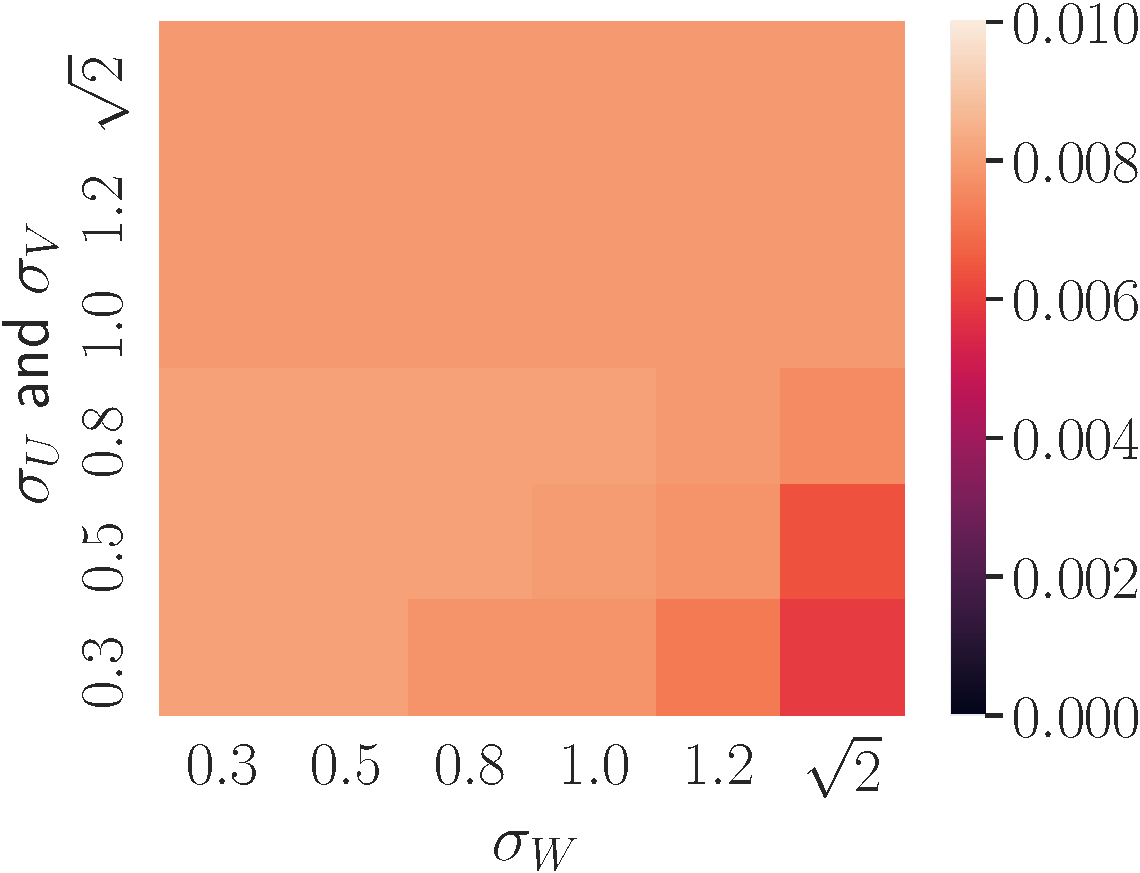}
    }\\
    \subfigure[OverRec on Toys.]{
    \includegraphics[width=0.144\linewidth]{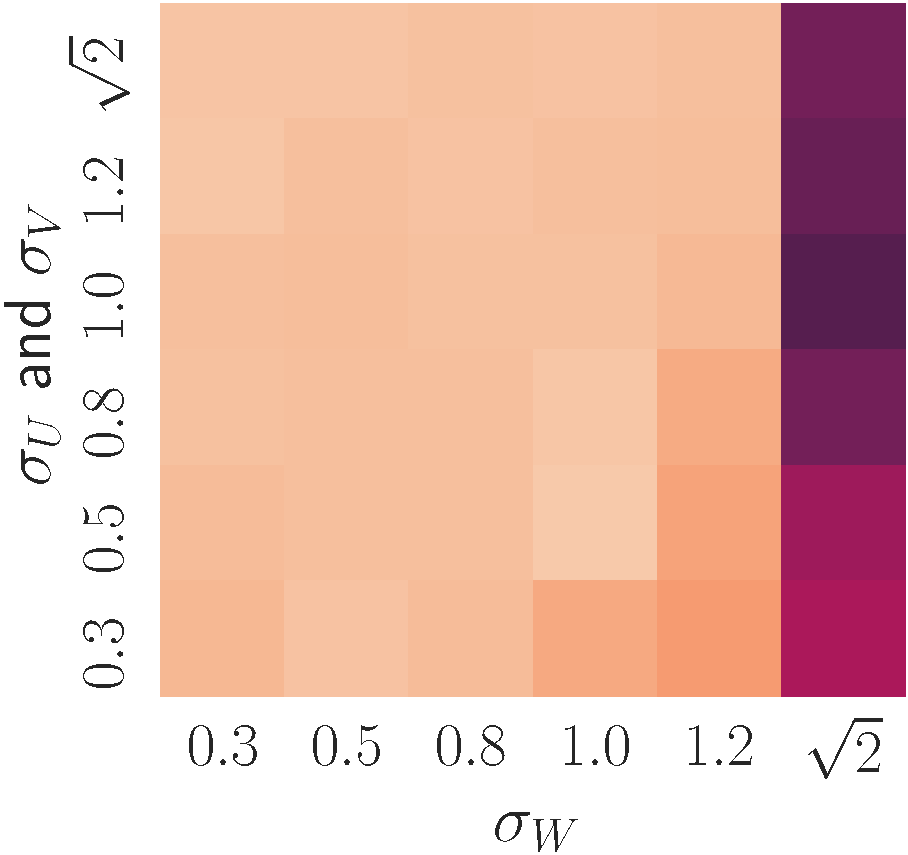}
    }
    \subfigure[NNGP-RNN on Toys.]{
    \includegraphics[width=0.179\linewidth]{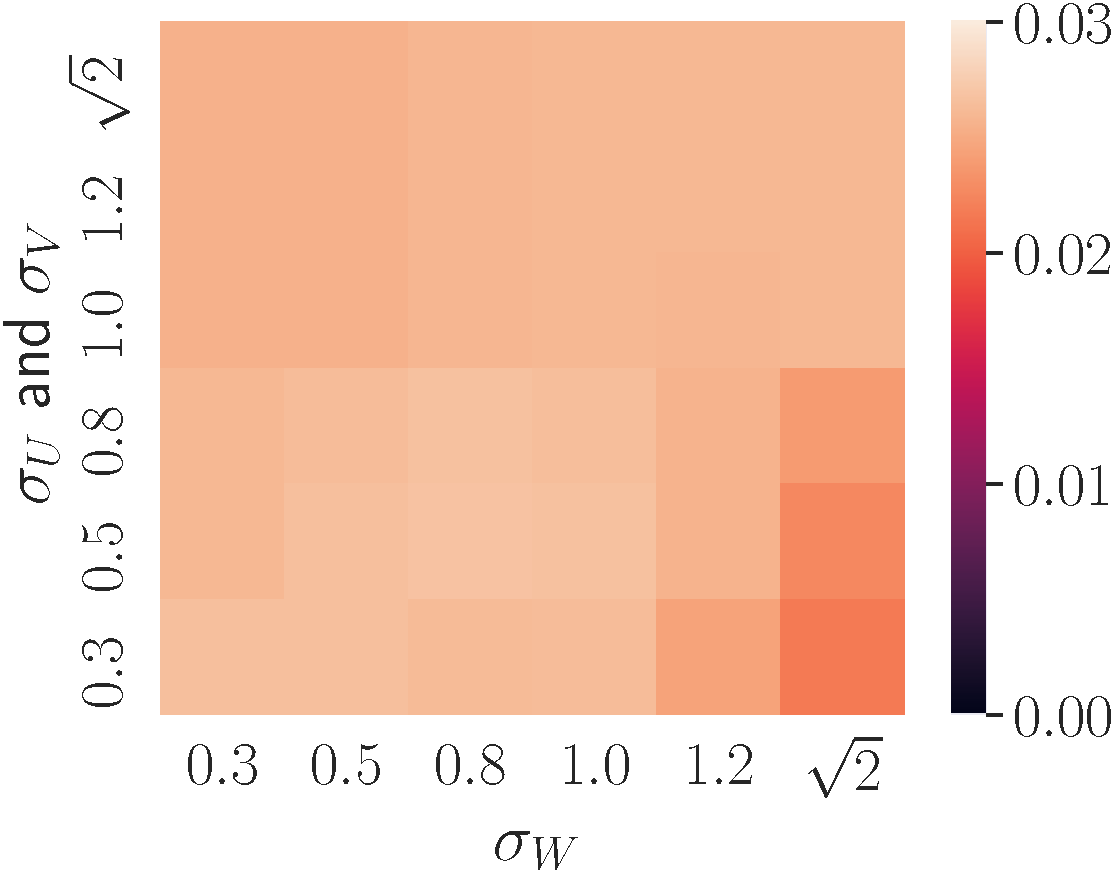}
    }
    \subfigure[OverRec on ML-1M.]{
    \includegraphics[width=0.146\linewidth]{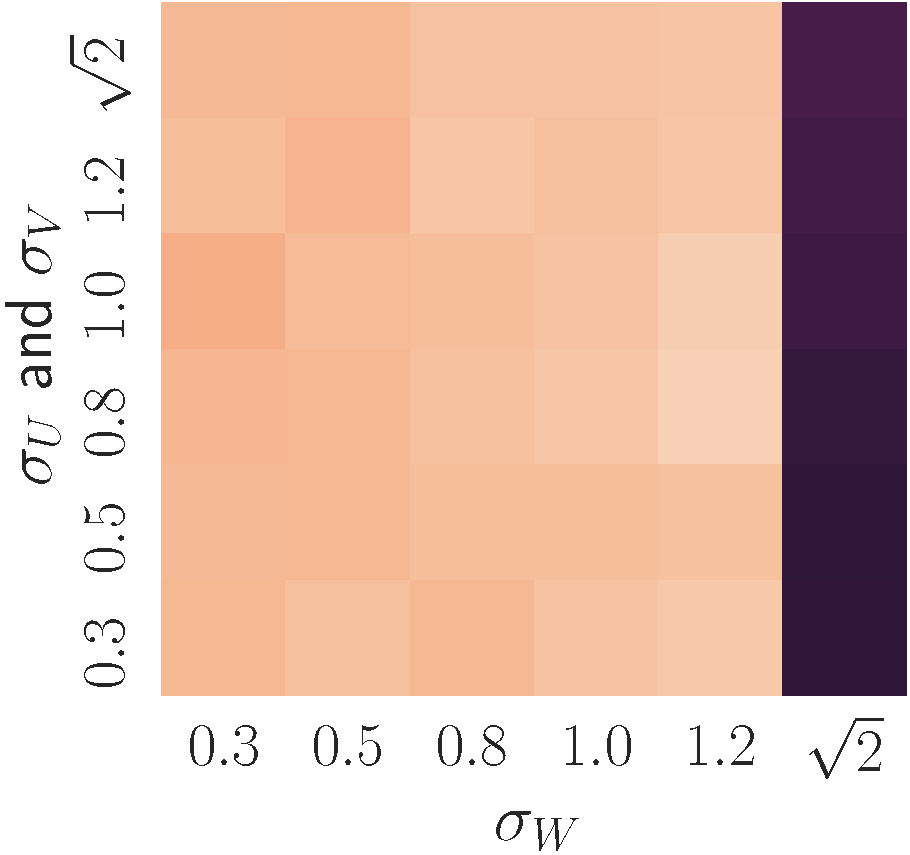}
    }
    \subfigure[NNGP-RNN on ML-1M.]{
    \includegraphics[width=0.18\linewidth]{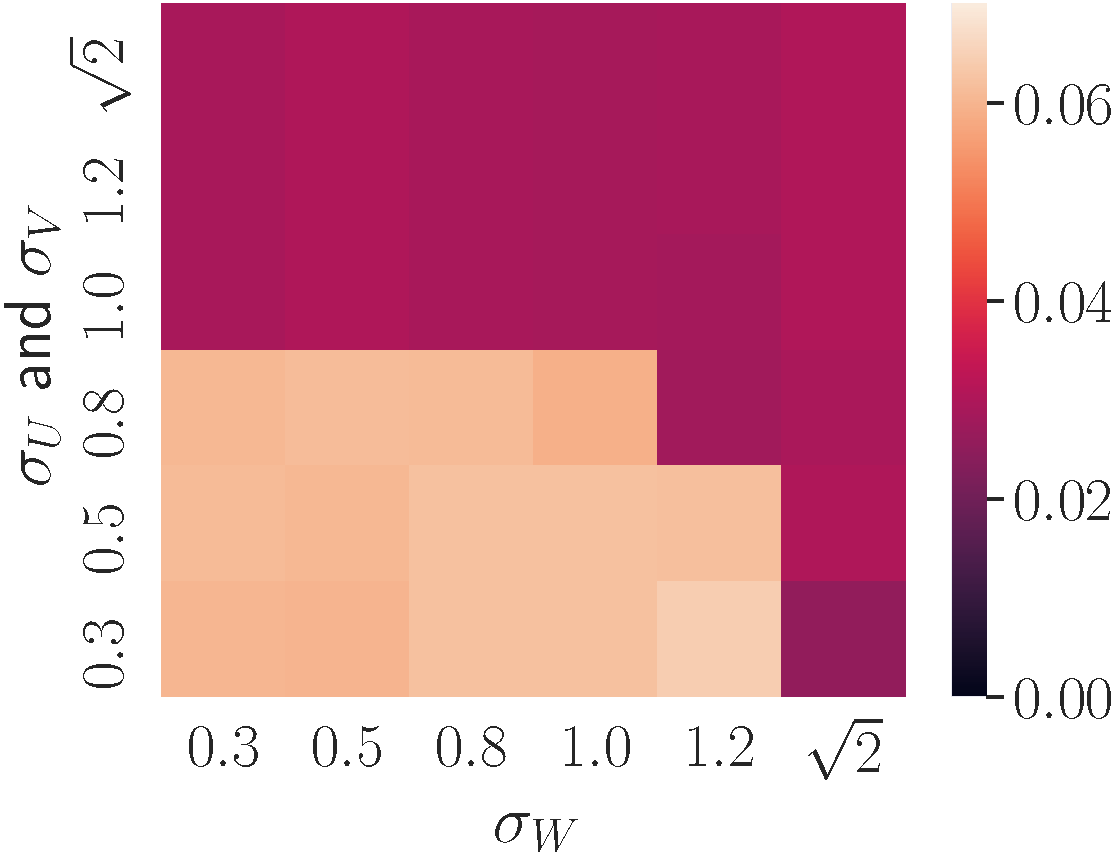}
    }
    \caption{MRR@5 of OverRec and the NNGP-RNN with different variances.}
\label{fig:param}
\end{figure*}

\subsection{Comparisons with $k$NN Methods}
\label{sec:rq-knn}
In this experiment, two major components in OverRec will be evaluated compared with the other two nearest neighbours-based methods: S$k$NN and STAN. In addition, the variant called NNGP-RNN is also evaluated here, which substitutes the RNTK in OverRec with NNGP-RNN. While for the prediction function, the Weighted Y from OverRec in Equation (\ref{eq:weighted_y}), the Equal Item from S$k$NN that considers all items in a sequence equally, and the Weighted Item from STAN that considers items with the time factors, are evaluated.

\subsubsection{Different Sequence Similarity Measurement}
For $k$NN-based methods, an important module is the similarity measurement function. In the proposed OverRec, this function is the RNTK. During the derivation of the RNTK, another kernel function NNGP-RNN is also calculated from Equation (\ref{eq:nngp-rnn}) and Algorithm~\ref{alg:nngp-rnn}, corresponding to the initialisation of an infinite-width RNN. Meanwhile, the previous methods such as S$k$NN and STAN will also be included.

From the results on Beauty, Clothing, Toys and ML-1M in Fig.~\ref{fig:knn}, it can be seen that in the left column using Weight Y as prediction function, the RNTK in OverRec consistently has the highest performance. NNGP-RNN has the second highest scores, which indicates that the neural network function at initialisation stage can perform a reasonable feature encoding. This is very similar to a random feature model~\cite{nngp,ntk-wide}. The worst performance comes from S$k$NN, which simply considers two sequences with more co-occurrence items as more semantically similar sequences. For the STAN method, which includes a weight factor defined by time in addition to S$k$NN, it achieves a higher result than S$k$NN, yet not comparable to OverRec and NNGP-RNN.

By using Weighted Y for prediction function, it requires a more accurate measurement of the semantic similarity of sequences because only the target item of the similar sequence will be used rather than all the items in the similar sequence. From this point of view, the higher the performance is, the better the sequence similarity is measured by the function. Therefore, it can be concluded that RNTK is the best sequence similarity measurement among all of these methods. NNGP-RNN has the second best ability in this measurement. While S$k$NN and STAN cannot sufficiently capture the semantic similarity of sequences by just using the item co-occurrence.

\subsubsection{Different Prediction Function}
For $k$NN methods, there is another important module, the prediction function. In OverRec, the Weighted Y from Equation (\ref{eq:weighted_y}) is used, only considering the target items from similar sequences. From S$k$NN, it includes all the items from the similar sequences for prediction. Since the choice of these items are with equal weight, it is referred to as Equal Item in this experiment. While for STAN, it proposes to assign different weights to the items from similar sequences based on the occurring time of these items. Therefore, this prediction function from STAN is referred to as Weighted Item.

From the results in Fig.~\ref{fig:knn}, it can be seen that for both OverRec and the NNGP-RNN, Weighted Y can always give out the highest performance, followed by Weighted Item. The Equal Item is the lowest across all methods in this experiment. According to the analysis in the accuracy of sequence similarity measurement, when the measurement is accurate such as RNTK and NNGP-RNN, Weighted Y is a better prediction because there is no noise from other items in these similar sequences. While the relatively higher performance of Weighted Item than Equal Item indicates that items at different time can have a different impact on the prediction. While for S$k$NN, Weighted Y is not a good choice for prediction since the similarity measurement from S$k$NN is not accurate and solely relying on the target item of not so similar sequences cannot generate a good prediction. For STAN, the performance of Weighted Y and Weighted Item is relatively similar.

\subsection{Parameter Sensitivity}
\label{sec:rq-param}
In this experiment, the parameter sensitivity of the variances in Equation (\ref{eq:init}) is investigated. The variances are chosen from $\{0.3,0.5,0.8,1.0,1.2,\sqrt{2}\}$ for $\sigma_W,\sigma_U$ and $\sigma_V$. Fig.~\ref{fig:param} demonstrates both OverRec and NNGP-RNN results of MRR@5.

From the figures, it is clear that for most of the parameter choices, the performance is stable for both OverRec and NNGP-RNN. For OverRec using RNTK, when $\sigma_W$ is increased, the performance will drop and until it reaches $\sqrt{2}$, the model becomes untrainable. While for NNGP-RNN, the worst case is higher than OverRec, yet it cannot achieve the same highest performance as OverRec. For example, when $\sigma_W=\sqrt{2}$, OverRec will generally fail to have a reasonable performance while NNGP-RNN can still perform well. This could be because at initialisation, the equivalent random neural network can still have the ability of feature encoding.

\section{Related Work}

\subsection{Sequential Recommendation}
The sequential recommendation task requires the model to perform next item prediction with sequential patterns. The main technique is sequential encoding~\cite{gru4rec,caser,sasrec,bert4rec,s3rec,safm,mminforec,agc,sslsb,npoi,fgnn,fgnnj,gag,posrec,causalrec}, which generally consists of recurrent neural networks and attention. GRU4Rec~\cite{gru4rec} is the first recurrent model using GRU. As the attention mechanism demonstrates more potentials, various models with attention are proposed such as SASRec~\cite{sasrec}, BERT4Rec~\cite{bert4rec}, $\text{S}^3$Rec~\cite{s3rec}, MMInfoRec~\cite{mminforec} and DuoRec~\cite{duorec}. The proposed OverRec is different from these methods that there is no explicit encoder in OverRec. Existing models exploit finite neural networks to encode user sequences for similarity calculation while OverRec has a kernel function to measure the sequence similarity with the properties of infinite-width neural networks.

Additionally, there are nearest neighbours-based methods. S$k$NN~\cite{sknn} and its extension VS$k$NN~\cite{vsknn} are based on the cosine similarity of the sequences. Using the time factor, STAN~\cite{stan} and its extension VSTAN~\cite{vstan} apply different weights by time.

\subsection{Infinite-Width Neural Networks}
Overparameterisation indicates training a neural network with the number of parameters larger than the number of training data points. Overparameterised models not only provide a better generalisation performance, but also give out theoretical insights in neural networks~\cite{over1,over2,ddd,dd}. Recent researches on traditional recommender systems empirically observe merits of overparameterisation for non-negative matrix factorisation with double descent~\cite{overnnmf} and normal matrix factorisation with acceleration~\cite{overmf}.
For the initialisation stage, Neal~\cite{nngp1} and Williams~\cite{nngp2} prove that the output of a two-layer fully connected network converges to a zero-mean Gaussian process. Recently, Lee et al.~\cite{nngp} show that the output of a wide enough fully connected network with any depth also converges to a zero-mean Gaussian process, called NNGP. Furthermore, Yang demonstrates that the output of most standard structures in deep learning such as a recurrent layer and convolutional layer, converges to NNGP~\cite{tp1gp4a}.

For the training stage, based on the NNGP results at initialisation, recent work finds out that infinite-width fully connected neural networks trained with stochastic gradient descent evolve according to an NTK~\cite{ntk,ntk-wide,ntk-exact}. When the training objective is mean square error, the infinite-width network behaves as a kernel ridge regression. In addition, other standard structures in deep learning have also been proved to have corresponding NTKs, CNTK for convolutional networks~\cite{ntk-exact}, GNTK for graph neural networks~\cite{gntk}, NTK for attention~\cite{antk} and RNTK for recurrent neural networks~\cite{rntk,rntkuci}. Yang and Littwin prove that the training dynamics of most structures in deep learning also correspond to NTK~\cite{tp2ntk4a}.

\section{Conclusion}
In this paper, the overparameterisation regime of sequential recommendation is empirically studied. Under this situation, a \textit{double ascent} phenomenon is observed in the test ranking performance, which is different from the traditional understanding of overfitting for large models. With this observation, an assumption is raised that if an infinite-width neural network model is built, the performance will be highly likely reach the limit of the specific neural network structure. For sequential recommendation, a novel OverRec model is proposed, in which the RNTK is developed to measure the similarity of sequences theoretically equivalent to a well-trained infinite-width RNN. We prove that the RNTKs for tied and untied input-output embeddings are the same. Extensive experiments are conducted on four datasets. Experimental results verify the superiority and the effectiveness of OverRec.

\section{Acknowledgments}
The work was supported by Australian Research Council (CE200100025, DP190102353, LP210300584, DP190101985, FT210100624).

\bibliographystyle{IEEEtranS}
\bibliography{IEEEabrv,icdm}

\end{document}